\documentclass[twocolumn,floatfix,aps,showpacs,showkeys]{revtex4-1}
\usepackage{epsfig,subfigure,latexsym}
\newcommand {\bea}{\begin{eqnarray}}
\newcommand {\eea}{\end{eqnarray}}
\newcommand {\be}{\begin{equation}}
\newcommand {\ee}{\end{equation}}

\usepackage{graphicx}
\usepackage{amssymb,amsmath}
\usepackage[normalem]{ulem}
\usepackage{rotating}
\usepackage[usenames]{color}

\begin{document}
\def\({\left(}
\def\){\right)}
\def\[{\left[}
\def\]{\right]}

\title{Correlation of the neutron star crust-core properties with the slope
of the symmetry energy and the lead skin thickness}

\author{H. Pais$^1$}
\author{A. Sulaksono$^2$} 
\author{B. K. Agrawal$^3$}
\author{C. Provid{\^e}ncia$^1$}

\affiliation{$^1$CFisUC, Department of Physics, University of Coimbra, 3004-516 Coimbra, Portugal.\\
$^2$Departemen Fisika, FMIPA, Universitas Indonesia, Depok, 16424,
Indonesia.  \\
$^3$Saha Institute of Nuclear Physics, Kolkata - 700064, India.
}

\begin{abstract}

The correlations of the crust-core transition density and pressure
in neutron stars with the slope of the symmetry energy and the
neutron skin thickness are investigated, using different families
of relativistic mean field  parametrizations with constant couplings
and non-linear terms mixing the $\sigma$, $\omega$ and $\rho$-meson
fields. It is shown that the modification of the density  dependence
of the symmetry energy, involving the $\sigma$ or the $\omega$ meson,
gives rise to different behaviors: the effect of the $\omega$-meson may
also be reproduced within non-relativistic phenomenological models,
while the effect of the $\sigma$-meson is essentially relativistic.
Depending on the parametrization with $\sigma-\rho$ or $\omega-\rho$
mixing terms, different values of the slope of the symmetry energy
at saturation must be considered in order to obtain a neutron matter
equation of state compatible with results from chiral effective field
theory. This difference leads to  different pressures at the
crust-core transition density. A  linear correlation between the
transition density and the symmetry energy slope or the neutron skin
thickness of the $^{208}$Pb nucleus  is obtained, only when the $\omega$-meson is used
to describe the density dependence of the symmetry energy. A comparison
is made between  the  crust-core transition properties of neutron stars
obtained by three different methods, the Relativistic Random Phase
Approximation (RRPA), the Vlasov equation and Thermodynamical method. It is shown that the RRPA and the Vlasov
methods predict similar transition densities for $pne$ $\beta$-equilibrium
stellar matter.

\end{abstract} 

\maketitle
 \section{INTRODUCTION}
\label{sec_intro}

Neutron stars (NS), with their extreme properties, like very high densities
and pressures, are an obvious laboratory to study nuclear physics, as
they are a window into the microscopic  properties of nuclear matter
at extreme isospin asymmetries \cite{Steiner-05}.
The properties of asymmetric nuclear systems have also been studied in
terrestrial laboratories for the past years \cite{Li08,Tsang-12}, but many
aspects have still to rely on theoretical models.  Constraints on the
behaviour of the symmetry energy above nuclear saturation density have
been coming from experiments  with new neutron-rich radioactive beams,
and in relativistic heavy-ion collisions, giant monopole resonances
\cite{Garg-07}, isobaric analogue states \cite{Danielewicz09} or
meson production (pions \cite{Li05}, kaons \cite{Fuchs06}) in heavy
ion collisions.

Correlations between different quantities in the bulk matter and finite
nuclei were established, like the correlation between the  pressure of
neutron matter at $\rho=0.1$ fm$^{-3}$ and the neutron skin thickness
of $^{208}$Pb \cite{Brown00,Typel01}, the correlation between the
crust-core transition density and the neutron skin thickness of $^{208}$Pb
\cite{Horowitz01}, the correlation between the slope $L$ and the curvature
$K_{sym}$ of the symmetry energy with the neutron skin thickness and the
crust-core transition density in compact stars \cite{Vidana09}. These
correlations, together with terrestrial and observational constraints,
will allow the construction of appropriate equations of state (EoS).

In Ref. \cite{Link-99}, it was shown that the glitches of Vela, which
are thought to  occur due to angular momentum transfer between the crust
and the core, could be explained if at least  1.4\% of the total moment
of inertia of the star resides in the inner crust.  Moreover, the same
authors also showed that the crustal moment of inertia is sensitive
to the pressure at the crust-core interface. Later, this mechanism was
questioned because neutron entrainement would require that the inner
crust contributes at least with  7\% of the total star moment of inertia
\cite{Chamel-13}. Entrainment seems to indicate that the crust is not
enough to account for the observed glitches \cite{Andersson-12}. Recently,
however, it was argued that uncertainties on the crust EoS are still large
and the mechanism of glitches may be totally explained by the crust,
if an appropriate EoS is considered, e.g. an EoS that predicts a large
transition pressure \cite{Piekarewicz-14}. This was possible by employing a
family of EoS where the density dependence of the symmetry energy was
accounted for, including in the Lagrangian density a term that mixes
the $\rho$ and the $\omega$-mesons.

Non-linear meson terms have been included in the Lagrangian
formulation of relativistic mean-field (RMF) models in order to modify the density dependence of the EoS, both
isoscalar and the isovector channels \cite{Mueller-96,Horowitz01}.  A
different approach has been considered in \cite{Typel-99}, where
non-linear terms were avoided at the expense of the inclusion of density
dependent couplings in the Lagrangian density. In \cite{Agrawal-10},
the  density dependence of the symmetry energy was described within an
extended RMF  model, including both self and mixed interaction terms
involving the scalar-isoscalar, vector-isoscalar  and vector-isovector
mesons up to the quartic order. The parameters of the models were
fitted to nuclear properties and the neutron thickness of the
$^{208}$Pb was allowed to vary in the range $\sim 0.20-0.24$ fm.

An expansion of the energy density of a system of nucleons described
within RMF   in powers of the Fermi momentum shows that the $\sigma$-meson
plays a special role in RMF models, giving rise to terms similar
to many-body repulsive terms in non-relativistic models \cite{sw,
ProvidenciaC-06a}. In fact, saturation is attained in RMF due to the
quenching of the $\sigma$-meson with density, while  non-relativistic
models have to introduce three-body repulsive interactions in order to
describe saturation correctly. In Ref. \cite{ASR2012}, the high density
EoS of nuclear matter was modified using a mixed $\sigma$-$\omega$ term,
and it has been shown that improvements were attained in the description
of the binding energy systematics and the EoS for the dilute
neutron matter, with respect to a simple quartic $\omega$ term. In the
present study, we will investigate the effect of  modificating the density
dependence of the symmetry energy using a mixed $\sigma$-$\rho$ term,
instead of a $\omega$-$\rho$ as in \cite{Piekarewicz-14}. $\sigma$-$\rho$
mixed terms were first included in \cite{Horowitz01}, where a quartic term
has been introduced in the Lagrangian density. In our study we consider,
instead, a third order term $\sim \sigma\rho^2$.

In this work, we investigate the correlation between the transition
density and pressure from the the inner crust to the core of neutron
stars. We use three different methods, the RRPA method \cite{AT06},
the Vlasov formalism \cite{Pais10}, and the thermodynamical method
\cite{ASR2012,Avancini-10,Ducoin-08,Ducoin11}, at zero temperature
and for $\beta-$equilibrium matter. We also investigate  the effect
of the contribution of the electrons, the Coulomb interaction and
the non-linear $\omega-\rho$ coupling term. The paper is organized as
follows. In Secs. \ref{sec_eos} and \ref{sec_formalism}, we introduce
the formalism used in this study, in Sec. \ref{sec_RaD}, we present and
discuss the results obtained and finally, in Sec. \ref{sec_conclu},
some conclusions are drawn.

\section{MODEL FOR NEUTRON STAR MATTER}
\label{sec_eos}

We use the relativistic non-linear Walecka model (NLWM) \cite{Walecka-74} in the mean-field approximation to study asymmetric nuclear and stellar matter at zero temperature.
We consider a system of baryons with mass $M$, interacting with
and through an isoscalar-scalar field $\sigma$ with mass $m_\sigma$, an
isoscalar-vector field $\omega^{\mu}$, with mass $m_\omega$, and an
isovector-vector field $\boldsymbol{\rho}^{\mu}$, with mass $m_\rho$. When
describing $npe$ matter, we also include a system of electrons with
mass $m_e$. Protons and electrons interact through the
electromagnetic field $A^{\mu}$. The Lagrangian density reads \cite{ASR2012}:
 \begin{equation}
\label{eq:lden}
{\cal L}= {\cal L}_{NM}+{\cal L}_e+{\cal L_{\sigma}} + {\cal L_{\omega}} + {\cal
L_{\mathbf{\rho}}} + {\cal L_{\sigma\omega{\rho}}}+{\cal L}_A. 
\end{equation}
Here, the Lagrangian ${\cal L}_{NM}$ describes the linear interactions of the
nucleons through the mesons exchange. The explicit form of   ${\cal L}_{NM}$ is

\begin{widetext}
\begin{eqnarray}
\label{eq:lbm}
{\cal L}_{NM} &=& \sum_{N=n,p}
\overline{\Psi}_{N}[i\gamma^{\mu}\partial_{\mu}-(M-g_{\sigma}
\sigma)-(g_\omega \gamma^{\mu} \omega_{\mu} \nonumber +\frac{1}{2}g_{\mathbf{\rho}}\gamma^{\mu}{\boldsymbol{\tau \cdot\rho}}_{\mu})]\Psi_{N}, 
\end{eqnarray}
\end{widetext}
where the sum is taken over the neutrons and protons, and 
$\boldsymbol{\tau}$ are the isospin matrices. 
The electron Lagrangian is given by
\begin{equation}
{\cal L}_e=\bar \psi_e\left[\gamma_\mu\left(i\partial^{\mu} + e A^{\mu}\right)
-m_e\right]\psi_e ,
\end{equation}
and
\begin{equation}
{\cal L}_A=-\frac{1}{4}F_{\mu\nu}F^{\mu\nu}
\end{equation}
with $F_{\mu\nu}=\partial_{\mu}A_{\nu}-\partial_{\nu}A_{\mu}$.
The Lagrangian densities describing the free mesons and
self interactions for $\sigma$, $\omega$, and $\rho$ mesons, respectively, can be
written as
\begin{equation}
\label{eq:lsig}
{\cal L_{\sigma}} =
\frac{1}{2}(\partial_{\mu}\sigma\partial^{\mu}\sigma-m_{\sigma}^2\sigma^2)
-\frac{{\kappa_3}}{6M}
g_{\sigma}m_{\sigma}^2\sigma^3-\frac{{\kappa_4}}{24M^2}g_{\sigma}^2 m_{\sigma}^2\sigma^4,
\end{equation}
\begin{equation}
\label{eq:lome}
{\cal L_{\omega}} =
-\frac{1}{4}\omega_{\mu\nu}\omega^{\mu\nu}+\frac{1}{2}m_\omega^2\omega_{\mu}\omega^{\mu}+\frac{1}{24}\zeta_0 g_\omega^{2}(\omega_{\mu}\omega^{\mu})^{2},
\end{equation}
\begin{equation}
\label{eq:lrho}
{\cal L_{\mathbf{\rho}}} =
-\frac{1}{4}{\boldsymbol{\rho}}_{\mu\nu}\cdot {\boldsymbol{\rho}}^{\mu\nu}+\frac{1}{2}m_{\rho}^2\boldsymbol{\rho}_{\mu}\cdot\boldsymbol{\rho}^{\mu}.
\end{equation}
The $\omega^{\mu\nu}$, $\boldsymbol{\rho}^{\mu\nu}$ are antisymmetric field tensors
corresponding to the $\omega$ and $\rho$ mesons. They are defined as
$\omega^{\mu\nu}=\partial^{\mu}\omega^{\nu}-\partial^{\nu}\omega^{\mu}$
and $\boldsymbol{\rho}^{\mu\nu}=\partial^{\mu}\boldsymbol{\rho}^{\nu}-
\partial^{\nu}\boldsymbol{\rho}^{\mu}-g_\rho\left(\boldsymbol{\rho}^\mu\times\boldsymbol{\rho}^\nu\right)$.  The mixing nonlinear $\sigma, \omega$, and $\rho$ mesons are described by ${\cal L_{\sigma\omega\rho}}$,

 \begin{eqnarray}
\label{eq:lnon-lin}
{\cal L_{\sigma\omega\rho}} & =&
\frac{\eta_1}{2M}g_{\sigma}m_\omega^2\sigma\omega_{\mu}\omega^{\mu}+ 
\frac{\eta_2}{4M^2}g_{\sigma}^2 m_\omega^2\sigma^2\omega_{\mu}\omega^{\mu} \nonumber \\
&+&\frac{\eta_{\rho}}{2M}g_{\sigma}m_{\rho }^{2}\sigma\boldsymbol{\rho}_{\mu}\cdot\boldsymbol{\rho}^{\mu} 
+\frac{\eta_{1\rho}}{4M^2}g_{\sigma}^2m_{\rho }^{2}\sigma^2\boldsymbol{\rho}_{\mu}\cdot\boldsymbol{\rho}^{\mu} \nonumber \\
&+&\frac{\eta_{2\rho}}{4M^2}g_\omega^2m_{\rho
}^{2}\omega_{\mu}\omega^{\mu}\boldsymbol{\rho}_{\mu}\cdot\boldsymbol{\rho}^{\mu} .
\end{eqnarray}

Of  particular interest in the present work are the cross-coupling terms
involving the $\rho$ meson field, which contributes to the isovector part
of the effective Lagrangian density, in addition to the usual linear
couplings of the $\rho$  meson to the nucleons.  We shall
mainly focus on the lowest order $\sigma-\rho$ and $\omega-\rho$
cross-couplings  whose strengths  are determined by the values of
$\eta_{\rho}$ and $\eta_{2\rho}$.  The quartic order $\sigma-\rho$
cross-coupling strength $\eta_{1\rho}$ is set to zero.   The values of
$\eta_\rho$ or $\eta_{2\rho}$ can be appropriately adjusted to yield  wide
variations in the density dependence of the symmetry energy coefficient
and the neutron skin thickness in heavy nuclei  without affecting the
other properties of finite nuclei \cite{Furnstahl-02,Sil-05,Dhiman-07}.

To study the role of the $\sigma \text{-} \rho$ and $\omega \text{-} \rho$
mixing terms on the crust-core transition properties in neutron stars,
we use two different  families of RMF model $F_{ \rho}$ and  $F_{2 \rho}$
\cite{Agrawal-10,Alam-15}.  The isovector part of the Lagrangian density
for the $F_\rho (F_{2\rho})$ family is governed by the couplings $g_\rho$
and $\eta_\rho (\eta_{2\rho})$.  The coupling $\eta_{1\rho}$ is set
to be zero for both the families.  The different parametrizations  of
$F_\rho$($F_{2\rho}$) families are obtained by  varying appropriately the
values of $g_{\rho}$ and $\eta_\rho(\eta_{2\rho})$. For a given value of
$\eta_\rho(\eta_{2\rho})$, the value of $g_\rho$ is always adjusted to
yield appropriate binding energy for the $^{208}$Pb nucleus.  Once the
values of  $g_\rho$ and $\eta_\rho$ or $\eta_{2\rho}$ are known, the
properties of the nuclear matter and the finite nuclei can be computed.
The values of $\eta_\rho$ and $\eta_{2\rho}$ are varied in the range of 
$0- 12$ and $0 - 60$, respectively.  

In Table \ref{tab1}, we present the parameters that differ among the models as considered in the present work. The remaining parameters, which correspond to the isoscalar part of the Lagrangian density, and the masses
of the $\sigma$, $\omega$ and $\rho$ mesons are kept fixed to those for
the BKA22 model \cite{Agrawal-10}.  The values of these parameters are $g_\sigma/(4\pi)=0.8462$, $g_\omega/(4\pi)=1.1089$, $k_3=1.55$, $k_4=2.13451$, $\zeta_0=5.8253$, $\eta_1=0.1555$, $\eta_2=0.0697$, and $\eta_{1\rho}=0$. The masses for the $\omega$, $\rho$  and $\sigma$ mesons are $m_\omega=782.5$ MeV,  $m_\rho=770$ MeV, and $m_\sigma=497.8578$ MeV. The nucleon mass is set to 939 MeV.
Table \ref{tab1} also shows the saturation properties, the neutron skin
thickness of $^{208}$Pb, and the crust-core transition densities,
$\rho_{trans}$,  calculated within the Vlasov formalism,
for $\beta-$equilibrium $pne$ matter at $T=0$ MeV, for all the models considered in the present work.

\begin{table}[t]
\caption{  \label{tab1}
The coupling  parameters $g_\rho$ and $\eta_\rho(\eta_{2\rho})$  for the
F$_\rho$ (F$_{2\rho}$)  families, and the corresponding values for the
symmetry energy coefficient, $E_{sym}$, and the symmetry energy slope
parameter, $L$, (in MeV), evaluated at the saturation density, $\rho_0$
=0.148 fm$^{-3}$. The values of the neutron skin thickness, $\Delta r_{\rm np}$,
(in fm), for the $^{208}$Pb nucleus, and the crust-core transition densities,
$\rho_{trans}$ (in fm$^{-3}$), calculated within the Vlasov formalism,
for $\beta-$equilibrium $pne$ matter at $T=0$ MeV, for all
the models are also listed.   The binding energy, $E/A$, is -16.08 MeV, the incompressibility, $K$, is 228.63 MeV and
the skewness coefficient, $Q_0$, is -285.03 MeV.}
\begin{ruledtabular}
\vspace{0.5cm}
\begin{tabular}{|ccccccc|}
Model &  $g_\rho$ &  $\eta_\rho$  & $E_{sym}$ & $L$ & $\Delta r_{\rm np}$& $\rho_{trans}$\\
    \hline
$F_\rho$-1 &  8.8614  & 0    &  36.4   & 108.9  & 0.280 & 0.0489 \\ 
$F_\rho$-2 & 11.1799  & 2.0 &  34.3   & 86.9   & 0.241  & 0.0570  \\
$F_\rho$-3 & 13.0335  & 4.0 &  33.4   & 79.1   & 0.219  & 0.0589 \\ 
$F_\rho$-4 & 14.6294  & 6.0 &  32.8   & 75.2   & 0.206  & 0.0591 \\ 
$F_\rho$-5 & 16.0494  & 8.0 &  32.4   & 72.8   & 0.194  & 0.0588  \\
$F_\rho$-6 & 17.3450  & 10.0 & 32.1   & 71.3   & 0.185 & 0.0585 \\
$F_\rho$-7 & 18.5388  & 12.0 & 31.9   & 70.1   & 0.179& 0.0582   \\
\hline
Model &$g_\rho$ & $\eta_{2\rho}$& $E_{sym}$ &$L$& $\Delta r_{\rm np}$& $\rho_{trans}$ \\
    \hline
$F_{2\rho}$-1 &  9.2225 & 2.5    & 35.6   & 97.3   & 0.269  & 0.0525 \\ 
$F_{2\rho}$-2 &  9.5656 & 5.0    & 35.0   & 88.5   & 0.259  & 0.0566 \\ 
$F_{2\rho}$-3 &  9.8898 &  7.5   & 34.5   & 81.7   & 0.250  & 0.0610 \\ 
$F_{2\rho}$-4 &  10.1964 & 10.0  & 34.1   & 76.2   & 0.242  & 0.0652 \\ 
$F_{2\rho}$-5 &  11.3135 & 20.0  & 32.8   & 62.5   & 0.215  & 0.0759 \\ 
$F_{2\rho}$-6 &  13.1771 & 40.0  & 31.3   & 50.9   & 0.178  & 0.0807  \\
$F_{2\rho}$-7 &  14.7328 & 60.0  & 30.4   & 46.0   & 0.152  & 0.0805 \\ 
  \end{tabular}
  \end{ruledtabular}
\end{table}

\section{METHODS TO CALCULATE NS CRUST-CORE PROPERTIES}
\label{sec_formalism}

In the present section,  we review the three methods that will be used
to determine the crust-core transition density: the Thermodynamical
method, the dynamical spinodal from the Vlasov equation and the RRPA formalism.

\subsection{Thermodynamical method}

For a system to be stable against small density fluctuations, the
thermodynamical method requires that  the conditions of
mechanical and chemical stabilities should be satisfied,
\cite{Kubis-04,Kubis-07,Lattimer-07} i.e.,
\begin{eqnarray}
\label{thermo1}
-\left (\frac{\partial P}{\partial v}\right )_{\hat{\mu}} > 0\\
-\left (\frac{\partial {\hat{\mu}}} {\partial q_c}\right )_{v} > 0
\label{thermo2}
\end{eqnarray}
Here $P = P_b + P_e$ is the total pressure of the  neutron, proton
and electron system, with the contributions $P_b$ and $P_e$ from baryons
and electrons, respectively. The $v$ and $q_c$ are the volume and charge
per baryon number. The $\hat{\mu}$ is the chemical potential defined as
$\hat{\mu} = {\mu}_n- {\mu}_p$.
Since the system under consideration is in $\beta$-equilibrium, which
implies $\hat{\mu} = \mu_e$, the electron pressure $P_e$  is a function
of $\hat{\mu}$. Eq. (\ref{thermo1}) becomes
\begin{equation}
-\left (\frac{\partial P_b}{\partial v}\right )_{\hat{\mu}} > 0\\.
\label{thermo3}
\end{equation}
Eqs. (\ref{thermo2}, \ref{thermo3})  can be expressed in terms of the energy
per nucleon $E_b(\rho, x_p)$ at a given density $\rho$ and proton fraction
$x_p$ as
\begin{equation}
-\left (\frac{\partial q_c}{\partial \hat{\mu}}\right)_v = \left (\frac{\partial^2
E_b(\rho, x_p)}{\partial x_p^2} \right )^{-1}+  \frac{\mu_e^2}{\pi^2 \hbar^3\rho}
\label{thermo4}
\end{equation}
\begin{widetext}
\begin{eqnarray}
-\left(\frac{\partial P_b}{\partial v}\right )_{\hat{\mu}}=\rho^2 \left[2\rho \frac{\partial
E_b(\rho, x_p)}{\partial \rho}+\rho^2 \frac{\partial^2 E_b(\rho, x_p)}{\partial^2 \rho} 
-\frac{ \frac{\partial^2 E_b(\rho, x_p)\rho^2 }{\partial \rho \partial x_p}} {\frac{\partial^2
E_b(\rho, x_p)}{\partial x^2_p}}\right] >0 
\label{thermo5}
\end{eqnarray}
\end{widetext}
 Eq. (\ref{thermo4}) is usually  valid, thus, the crust-core transition
density  is determined by using the inequality of Eq. (\ref{thermo5}).

\subsection{Vlasov method}

In Refs. \cite{Nielsen-91, Nielsen-93}, the collective modes in
cold nuclear matter were determined within the Vlasov equation,
based on the Walecka model \cite{Walecka-74}, and later also used in
Refs. \cite{ProvidenciaC-06a,Avancini-05}. We have extended this formalism to
include the mixing non-linear self-interactions of the mesons $\sigma,
\omega$ and $\rho$ in Ref. \cite{Pais-vlasov}.

The time evolution of the distribution functions, $f_i$, is described by the
Vlasov equation, which expresses the
conservation of the number of particles in phase space, and is, therefore, covariant.
At zero temperature, the state that minimizes the energy of asymmetric nuclear matter is characterized
by the Fermi momenta $P_{Fi} ,i = p, n$, $P_{Fe} = P_{Fp}$, and is described by the distribution function 
\begin{equation}
f_0({\boldsymbol r},{\boldsymbol p})=\mbox{diag}[\Theta(P_{Fp}^2-p^2),\Theta(P_{Fn}^2-p^2),\Theta(P_{Fe}^2-p^2)]
\end{equation}
 and by
the constant mesonic fields. 

Collective modes in the present approach correspond to small oscillations
around the equilibrium state. These small deviations are described by the
linearized equations of motion and collective modes are given as
solutions of those equations. 

The linearized Vlasov equations
for $\delta f_{i}$,
$$\frac{d\delta f_{i}}{d t}+ \{\delta f_{i}, h_{0 i}\}
 +\{f_{0 i},\delta h_{i} \}=0$$
 are equivalent to the following time-evolution equations:

\begin{eqnarray}
\label{eq:deltaf}
\frac{\partial S_{i}}{\partial t} + \{S_{i },h_{0i}\} = \delta h_{i}
&=& -g_\sigma\, \frac{ M^*}{\epsilon_{0}}\delta\phi 
- \frac{{\bf p} \cdot \delta \boldsymbol{\cal
V}_i}{\epsilon_{0}}  + \delta{\cal V}_{0i},\,\, \nonumber \\
i&=&p,n
\end{eqnarray}

\begin{equation}
  \label{eq:deltafe}
\frac{\partial S_{e}}{\partial t} + \{S_{e},h_{0e }\} = \delta h_{e}
= -e\left[ \delta{A}_{0} - \frac{{\bf p} \cdot \delta{\mathbf A}}{\epsilon_{0e}}\right],
\end{equation}
where
\begin{eqnarray*}
\delta f_i &=&\{S_i,f_{0i}\}\,  ,\\
\delta{\cal V}_{0i}&=&g_\omega \delta \omega_0 + \tau_i \frac{g_{\rho}}{2}\,
\delta \rho_0 + e\, \frac{1+\tau_{i}}{2}\,\delta A_0 ,\\ \nonumber
\delta \boldsymbol{\cal V}_i&=& g_\omega \delta \boldsymbol{\omega} + \tau_i
 \frac{g_{\rho}}{2} \,\delta \boldsymbol{\rho}+ e\, \frac{1+\tau_{i}}{2}
\delta {\mathbf A}\, , \\
 h_{0 i}&=&\epsilon_{0} +{\cal V}^{(0)}_{0 i}\,=\sqrt{p^2+M^{*2}}+{\cal V}^{(0)}_{0 i} \, , \\
h_{0 e}&=& \epsilon_{0 e}=\sqrt{p^2+m^{2}_e} \, ,
\end{eqnarray*}
which has only to be satisfied for $p=P_{Fi}$.

We are interested in the longitudinal modes, with wave vector ${\bf k}$ and frequency $\omega$, which are
described by the ansatz
\begin{equation}
\left(\begin{array}{c}
S_{j}({\bf r},{\bf p},t)  \\
\delta\phi  \\
\delta \zeta_0 \\ \delta \zeta_i
\end{array}  \right) =
\left(\begin{array}{c}
{\cal S}_{\omega}^j ({\rm cos}\theta) \\
\delta\phi_\omega \\
\delta \zeta_\omega^0\\ \delta \zeta_\omega^i
\end{array} \right) {\rm e}^{i(\omega t - {\bf k}\cdot
{\bf r})} \;  ,
\label{ansatz}
\end{equation}
where $j=p,\, n,\, e$; $\zeta=\omega,\, \rho,\, A$ represent the
vector-meson fields, and $\theta$ is the angle between ${\bf p}$
and ${\bf k}$. The wave vector of the excitation mode,  ${\bf k}$,
is identified with the momentum transferred to the system through
the process which gives rise to the excitation.

Replacing the ansatz (\ref{ansatz})
in Eqs. (\ref{eq:deltaf}) and  (\ref{eq:deltafe}), we get a set of equations for the fields and for the generating functions. The solutions of these equations form
a complete set of eigenmodes that may be used to construct a
general solution for an arbitrary longitudinal perturbation. They  lead to the following
matrix equation
\begin{equation}
\left(\begin{array}{ccc}
1+F^{pp}L_p&F^{pn}L_p&C_A^{pe}L_p\\
F^{\rm np}L_n&1+F^{nn}L_n&0\\
C_A^{ep}L_e&0&1-C_A^{ee}L_e\\
\end{array}\right)
\left(\begin{array}{c}
A_{\omega p} \\
A_{\omega n} \\
A_{\omega e}
\end{array}\right)=0. \label{matriz}
\end{equation}
with $A_{\omega i}=\int_{-1}^1xS_{\omega i}(x)dx$, $L_i=L(s_i)=2-s_i\ln((s_i+1)/(s_i-1))$, where $s_i=\omega/\omega_{0i}$, and $F^{ij}=C_s^{ij}-C_v^{ij}-C_{\rho}^{ij}-C_A^{ij}\delta_{ip}\delta_{ij}$. The coefficients $C_s^{ij}, C_v^{ij}$,  $C_{\rho}^{ij}$ and $C_A^{ij}$ are given in Ref. \cite{Pais-vlasov}.

At subsaturation densities, there are unstable modes identified by imaginary
frequencies. For these modes we define the growth rate $\Gamma=-i\omega$.
The region in ($\rho_p,\rho_n$) space for a given wave vector $\mathbf k$ and
temperature $T$,  limited by the surface $\omega=0$,
defines the dynamical spinodal surface.  
 In the  $k=0$ MeV limit, we recover the thermodynamic
spinodal, which is defined by the surface in the ($\rho_p,\, \rho_n,\,
T$) space for which the curvature matrix of the free energy density is
zero, i.e., has a zero eigenvalue. This relation has been
discussed in \cite{ProvidenciaC-06}.

\subsection{RRPA method}

The longitudinal dielectric function can be written as~\cite{Carriere03}
\be
\varepsilon_L = {\rm{det}} \[1-D_L(q) \Pi_L(q,q_0=0)\] . \label{eq:det}
\ee 
The uniform ground state system becomes unstable to small-amplitude
density fluctuations with momentum transfer $q$, when $\varepsilon_L$ $\le$ 0. 
Note that in Eq.~(\ref{eq:det}), $q_0$ is the time-component of the
four-momentum transfer $q^\mu=(q_0,\vec{q}\,)$ and $q=|\vec{q}\,|$.
The transition density, $\rho_t$, is the largest
density for which the above condition has a solution.
For matter that in general consists of protons, neutrons, electrons and muons, the 
longitudinal meson propagator is given by
\bea
D_L = \left( \begin{array} {ccccc}
d_g & d_g& 0 & -d_g & 0\\
d_g & d_g& 0 & -d_g & 0\\
0&0& -d_s & d_{sv\rho}^+ & d_{sv\rho}^-\\
-d_g&-d_g &  d_{sv\rho}^+ & d_{33} & d_{v\rho}^-\\
0 &0& d_{sv\rho}^- &  d_{v\rho}^- & d_{44}
\end{array}\right) ~,
\label{eq:longprop}
\eea
where $d_{sv\rho}^+ = - (d_{sv} +   d_{s\rho}$), $d_{sv\rho}^- =  -(d_{sv} -   d_{s\rho}$), $d_{v\rho}^- =  d_{v} -   d_{\rho}$, $d_{33} = d_g +  d_{v} + d_{\rho} + 2  d_{v\rho}$ and $d_{44} = d_{v} + d_{\rho} - 2  d_{v\rho}$. In this form, mixing propagators between isoscalar-scalar and isoscalar-vector ($d_{sv}$), isoscalar-vector and isovector-vector ($d_{v\rho}$), isoscalar-scalar and isovector-vector ($d_{s\rho}$) are present due to the mixing self-interaction nonlinear terms in RMF model, in addition to the standard  $\gamma$, $\omega$, $\sigma$ and $\rho$ propagators ($d_g$,  $d_{v}$, $d_{s}$ and  $d_{\rho}$). These propagators are determined from the quadratic fluctuations around the static solutions which are generated by the second derivatives of energy density (${\partial^2 \epsilon}/{\partial \phi_i \partial\phi_j}$), where $\phi_i$ and $\phi_j$ are the  involved meson fields.
The explicit forms of the $\sigma$, $\omega$, and $\rho$ propagators are
\begin{widetext}
\bea
d_s&=&\frac{g_{\sigma}^2 (q^2+m_\omega^{*~2})(q^2+m_\rho^{*~2})}{(q^2+m_\omega^{*~2})(q^2+m_\rho^{*~2})(q^2+m_\sigma^{*~2})+(\Pi_{\sigma \omega}^0)^2(q^2+m_\rho^{*~2})+(\Pi_{\sigma \rho}^0)^2 (q^2+m_\omega^{*~2})}~,\nonumber\\
d_v&=&\frac{g_{\omega}^2 (q^2+m_\sigma^{*~2})(q^2+m_\rho^{*~2})}{(q^2+m_\omega^{*~2})(q^2+m_\rho^{*~2})(q^2+m_\sigma^{*~2})+(\Pi_{\sigma \omega}^0)^2(q^2+m_\rho^{*~2})-(\Pi_{\omega \rho}^{00})^2 (q^2+m_\sigma^{*~2})}~,\nonumber\\
d_\rho&=&\frac{1/4 g_\rho^2 (q^2+m_\sigma^{*~2})(q^2+m_\omega^{*~2})}{(q^2+m_\omega^{*~2})(q^2+m_\rho^{*~2})(q^2+m_\sigma^{*~2})+(\Pi_{\sigma \rho}^0)^2(q^2+m_\omega^{*~2})-(\Pi_{\omega \rho}^{00})^2 (q^2+m_\sigma^{*~2})}~,\nonumber\\
\label{eq:d3}
\eea
\end{widetext}
and the meson mixing propagators take the following form
\bea
d_{sv}&=&\frac{g_{\sigma} g_{\omega} \Pi_{\omega \sigma}^{0}(q^2+m_\rho^{*~2})}{H(q,q_0=0)}~,\\
d_{s\rho}&=&\frac{1/2 g_\rho g_{\sigma} \Pi_{\sigma \rho}^{0}(q^2+m_\omega^{*~2})}{H(q,q_0=0)}~,\\
d_{v \rho}&=&\frac{1/2 g_\rho g_{\omega} \Pi_{\omega \rho}^{00}(q^2+m_\sigma^{*~2})}{H(q,q_0=0)}~,
\eea
where the explicit form of $H(q,q_0=0)$ can be written as
\bea
H(q,q_0=0)&=&(q^2+m_\omega^{*~2})(q^2+m_\rho^{*~2})(q^2+m_\sigma^{*~2})\nonumber\\&+&(\Pi_{\sigma \omega}^0)^2(q^2+m_\rho^{*~2})+(\Pi_{\sigma \rho}^0)^2 (q^2+m_\omega^{*~2})\nonumber\\&-&(\Pi_{\omega \rho}^{00})^2 (q^2+m_\sigma^{*~2})~,
\label{HABC}
\eea
and the meson effective masses in Eq.~(\ref{HABC}) are defined as \\
\begin{widetext}
\bea
m_\sigma^{*~2}&=& \frac{\partial^2 \epsilon}{\partial^2 \sigma}=m_\sigma^2+ \frac{g_\sigma m_\sigma^2 \kappa_3}{ M} \sigma+ \frac{g_\sigma^2 m_\sigma^2 \kappa_4}{2 M^2}  \sigma^2 -  \frac{g_\sigma^2 m_\omega^2 \eta_2}{2 M^2}  \omega_0^2 -  \frac{g_\sigma^2 m_\rho^2 \eta_{1\rho}}{2 M^2} \rho_0^2~,\\ 
m_\omega^{*~2}&=& -\frac{\partial^2 \epsilon}{\partial^2 \omega_0}=m_\omega^2+  \frac{g_\sigma m_\omega^2 \eta_1}{ M}\sigma+ \frac{g_\sigma^2 m_\omega^2 \eta_2}{2 M^2} \sigma^2 +  \frac{\zeta_0 g_\omega^2}{2 } \omega_0^2 +\frac{ g_\omega^2 m_\rho^2 \eta_{2\rho}}{2 M^2}    \rho_0^2~, \\
m_\rho^{*~2}&=& -\frac{\partial^2 \epsilon}{\partial^2 \rho_0}=m_\rho^2+\frac{ g_\sigma m_\rho^2 \eta_{\rho}}{M} \sigma + \frac{g_\sigma^2 m_\rho^2 \eta_{1\rho}}{2 M^2} \sigma^2 +  \frac{ g_\omega^2 m_\rho^2 \eta_{2\rho}}{2 M^2}  \omega_0^2~,
\label{eq:meseffmass}
\eea
\end{widetext}
while the polarization due to mesons mixing self-interaction nonlinear terms in Eq.~(\ref{eq:lnon-lin}) (Mix Polarizations) are
\bea
\Pi_{\sigma \omega}^0&=&-\frac{\partial^2 \epsilon}{\partial \sigma \partial \omega_0}=  \frac{g_\sigma m_\omega^2 \eta_1}{ M} \omega_0 +  \frac{g_\sigma^2 m_\omega^2 \eta_2}{ M^2} \sigma \omega_0~,\nonumber\\ \Pi_{\sigma \rho}^0&=&-\frac{\partial^2 \epsilon}{\partial \sigma \partial \rho_0}=  \frac{ g_\sigma m_\rho^2 \eta_{\rho}}{M} \rho_0 +  \frac{g_\sigma^2 m_\rho^2 \eta_{1\rho}}{ M^2} \sigma \rho_0~,\nonumber\\ \Pi_{\omega \rho}^{0 0}&=&\frac{\partial^2 \epsilon}{\partial \omega_0 \partial \rho_0}= - \frac{ g_\omega^2 m_\rho^2 \eta_{2\rho}}{ M^2} \omega_0 \rho_0 ~,
\label{eq:polar}
\eea
whereas the propagator of photon takes a standard form i.e.,  
\bea
d_g = \frac{e^2}{q^2} ~.
\label{eq:Coul}
\eea

The longitudinal polarization matrix given in Eq.~(\ref{eq:det}) reads
\bea  
\Pi_L = \left( \begin{array} {ccccc}
\Pi_{00}^e & 0 & 0 & 0&0\\
0& \Pi_{00}^{\mu}  & 0 & 0&0\\
0 &0& \Pi_{s}& \Pi_{m}^p &\Pi_{m}^n\\
0 &0& \Pi_{m}^p & \Pi_{00}^p & 0\\
0 &0& \Pi_{m}^n & 0 & \Pi_{00}^n
\end{array}\right) ~.
\label{eq:longpol}
\eea

The formulas for polarization elements in $\Pi_L$ are given in, e.g.,
Ref.~\cite{AT06}. 

For the families of models under study in the present article, the $\Pi_{\sigma \omega }^0$
contribution is present, but for the $F_{2\rho}$ family, the contribution from $\Pi_{\sigma
\rho}^0$ is zero, and for the $F_{ \rho}$ family, the $\Pi_{\omega \rho}^0$ contribution
becomes zero. In the crust-core region, usually the muons have not yet appeared, so that  $\Pi_{00}^{\mu}$ can be set to zero, and if we consider the case without electrons, then $\Pi_{00}^e$ is also set to zero.

\section{Results and discussion}
\label{sec_RaD}

\begin{figure}
\begin{tabular}{c}
\includegraphics[width=0.5\textwidth]{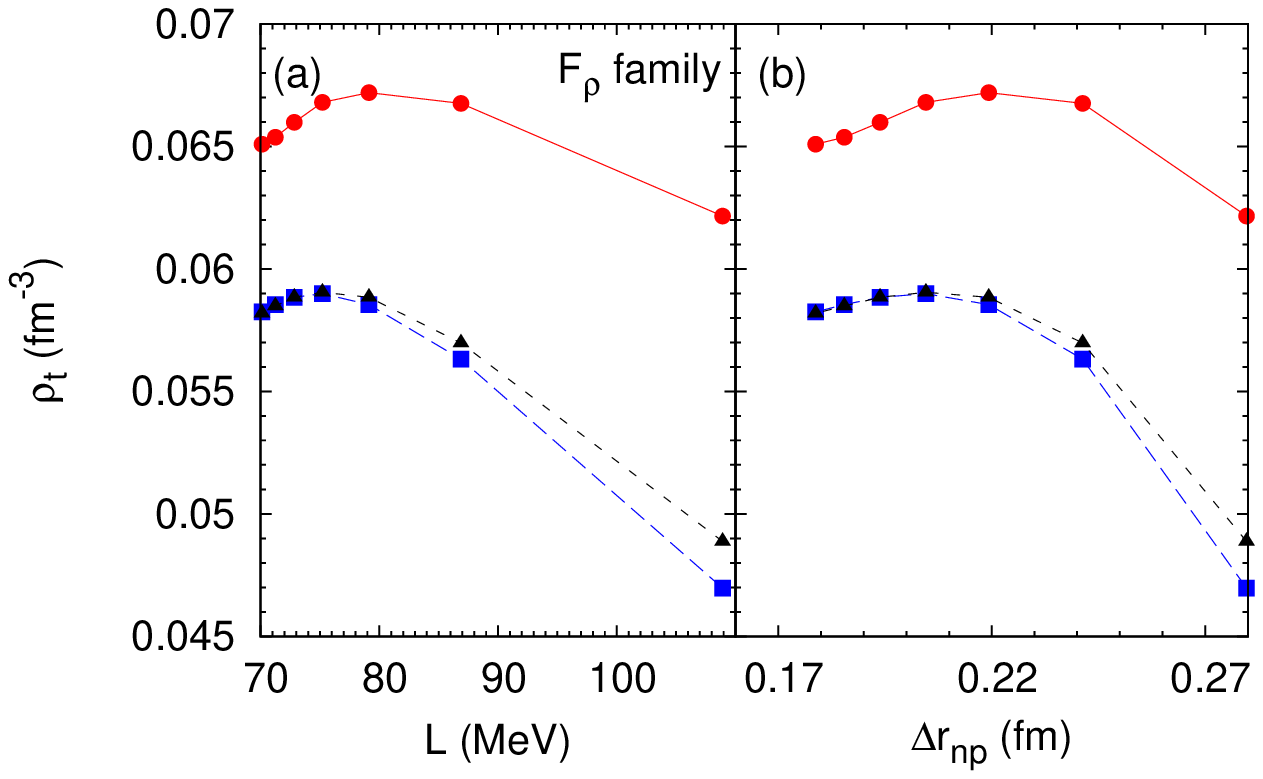}  \\
\includegraphics[width=0.5\textwidth]{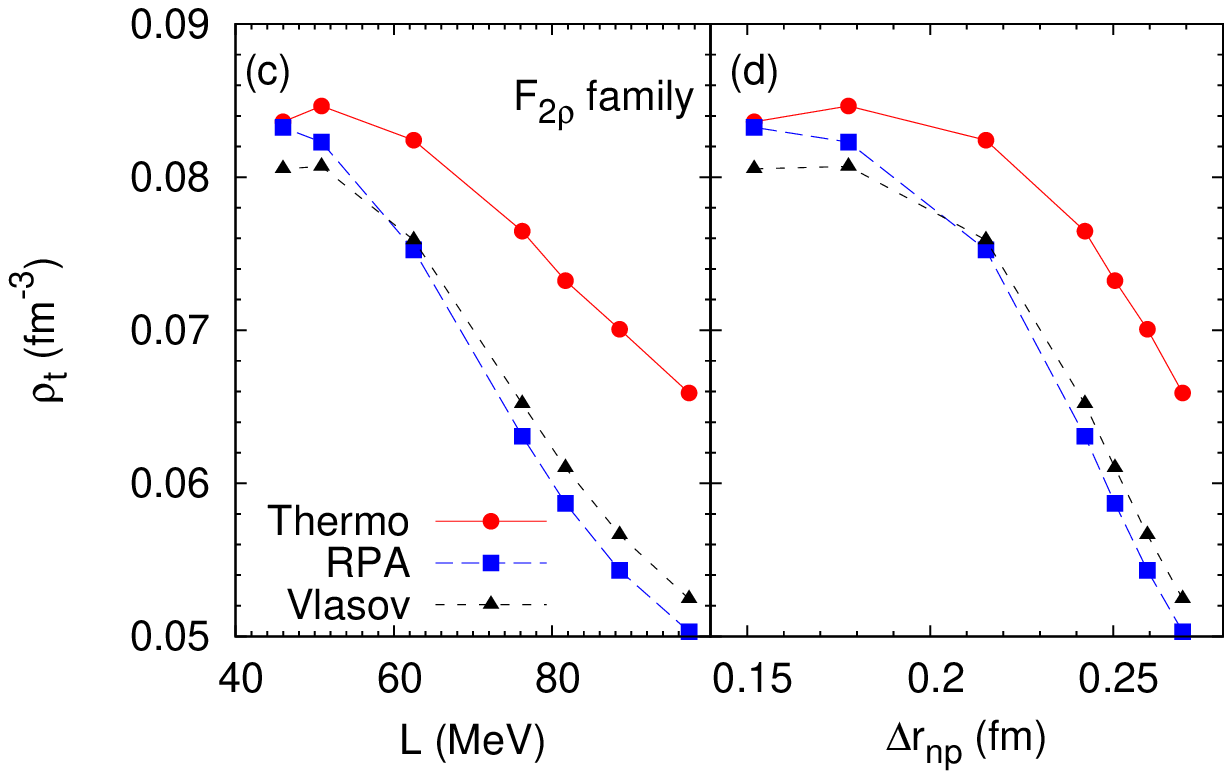}  
\end{tabular}
\caption{(Color online) Crust-core transition density, $\rho_t$,  for
  the F$_\rho$ ((a) and (b)) and F$_{2\rho}$ ((c) and (d)) families as a
  function of $L$ (left panels) and $\Delta r_{\rm np}$ (right panels)
from RRPA, Vlasov and thermodynamical methods. }
\label{Fig:1}
\end{figure}

In Figure ~\ref{Fig:1}, we plot the crust-core transition density,
$\rho_t$, as a function of the slope of the symmetry energy $L$ [(a) and (c)], 
and the neutron skin thickness $\Delta r_{\rm np}$ [(b) and (d)] for
the F$_\rho$ family (upper panel) and for the F$_{2\rho}$ family (lower panel). The slope of the symmetry energy is defined as $L=3\rho_0\left(\partial E_{sym}/\partial \rho\right)_{\rho=\rho_0}$, see, e.g., \cite{Vidana09}.
The  values of $\rho_t$ are calculated  for
the three different methods: RRPA, Vlasov, and
Thermodynamical. 
We can see that the transition densities calculated from both the RRPA
and Vlasov approaches agree quite well with each other for both families. 
These results are, therefore, compatible with previous studies
  executed within Skyrme interactions \cite{Ducoin-08a} or RMF models
  \cite{Horowitz-08}, where RRPA calculations and semiclassical calculations
  were compared.

Comparing the behavior of both
families, the F$_{2\rho}$ family shows, overall, higher transition
densities to uniform matter than the $F_\rho$  family. The correlation
between  $\rho_t$ and $L$ (or $\Delta r_{\rm np}$) depends significantly on the
kind of meson self-interaction mixing terms used in the RMF model. 
The $\rho_t$ evolution predicted by the $F_{\rho}$ family shows a linear correlation with $L$ and the neutron thickness in the low $L$ (or $\Delta r_{\rm np}$) region.
 This behavior differs from the one of the $F_{2 \rho}$ family, where the transition density always decreases with increasing $L$ (or $\Delta r_{\rm np}$), and from previous works  \cite{Brown00,Typel01,Horowitz01,Vidana09,Ducoin-08,Ducoin11}.

The density  dependence  of the symmetry energy is achieved through the
inclusion of a mixed term of the $\rho$-meson with the $\sigma$-meson
for $F_\rho$ and the $\omega$-meson for $F_{2\rho}$ families. At subsaturation,
the behavior of these two mesons, or, correspondingly, the scalar and
the nucleonic densities,  are quite different, in particular, the
$\sigma$ field, which is responsible for binding the matter,  is stronger
and increases much faster with density, for the values close to zero,
and slower above $\sim 0.08$ fm$^{-3}$. 
The families $F_\rho$ and $F_{2\rho}$ were built 
 by fitting the binding energy for the $^{208}$Pb nucleus for
 different values of the neutron skin thickness.
 Due to the different behavior of the
$\sigma$ and $\omega$ mesons, the first one determining the behavior
of the family $F_\rho$, and the second one the behavior of the family $F_{2\rho}$, for a given neutron skin thickness, the slope $L$ and the $g_\rho$
coupling for the parametrizations with the $\sigma-\rho$  term
(family $F_\rho$) is larger. For this same family, 
the transition density is lower and almost does not change below
$L=80$ MeV.
As we will also see, the pressure behaves differently.

\begin{table}
\caption{\label{tab2} The values of the total binding energy ($E$ ) in
MeV, charge radii ($r_c$), neutron radii ($r_n$) and $\Delta r_{\rm np}$
in fm for a few asymmetric spherical nuclei obtained for selected  parameterisations of the
$F_{\rho}$ and $F_{2\rho}$ family.}

 \begin{tabular}{|cc|ccccc|}
 \hline
 & &   Expt& $F_{\rho}-3$&  $F_{\rho}-7$ & $F_{2\rho}-5$& $F_{2\rho}-6$\\
\hline
$^{48}{\rm Ca}$&$E$&-416.00&   -415.75 &   -415.10&  -415.71&  -415.04\\
&$r_c$&           3.477  &  3.468 &   3.478& 3.466& 3.473 \\
&$r_n$&             &  3.575  &  3.561 & 3.571& 3.559 \\
&$\Delta r_{\rm np}$&&          0.201 &   0.178&  0.199& 0.179 \\
\hline
$^{132}{\rm Sn}$&$E$&1102.84  &  -1102.49 &  -1101.01 &   -1102.49&     -1101.07 \\
&$r_c$&            4.709   &  4.736 &  4.751 &  4.728&   4.739   \\
&$r_n$&               &    4.952  &4.924  &4.939 & 4.911 \\
&$\Delta r_{\rm np}$&   &      0.284 &  0.240 & 0.279& 0.239  \\
\hline
$^{208}{\rm Pb}$&$E$&-1636.43 & -1637.07  & -1637.07&  -1637.09 & -1637.03  \\
&$r_c$&              5.501  &  5.545&  5.559 & 5.537  &5.547\\
&$r_n$&                &  5.706 & 5.680  &5.694  &5.666\\
&$\Delta r_{\rm np}$&  & 0.219 &   0.178  & 0.215   &  0.178 \\
\hline
\end{tabular}
\end{table}

It is important to look into the  bulk properties of finite
nuclei obtained for each of the  families.  Table \ref{tab2}  contains
the total energy, the charge radii, the neutron radii and the neutron
skin thickness for three  isospin asymmetric spherical nuclei $^{48}$Ca,
$^{132}$Sn and $^{208}$Pb, for selected parametrizations of
the $F_\rho$  and $F_{2\rho}$ family. The parameter sets are so chosen
that they predict similar neutron skin thicknesses in $^{208}$Pb for
both the  families.  The differences between the charge radii for these
nuclei obtained for the two selected models within each family  are
equal or smaller than 0.015 fm for the family $F_\rho$ and 0.011 fm for
the family $F_{2\rho}$.  For similar values of $\Delta r_{\rm np}$, the
charge radii for the family $F_\rho$ are larger by  $\sim 0.01$fm than
the ones obtained for the family $F_{2\rho}$ .  These differences can be
levelled off by increasing the saturation density $\sim 0.001$fm$^{-3}$
in the case of $F_\rho$ family.  However, such fine tuning may not
explain the observed trend of the transition density that it  stays
almost unchanged below $\Delta r_{\rm np}=0.22$fm for the family $F_\rho$.
The values of  charge radii  for the $F_\rho$ and $F_{2\rho}$ families
are within the prediction of other RMF parametrizations that have been
tuned to nuclear properties \cite{Sugahara94,Todd05,Lalazissis-97} and are employed for the study of correlations of
$\Delta r_{\rm np}$ with various bulk properties of nuclear matter and the neutron stars  \cite{Horowitz01,Carriere03}.

\begin{figure*}
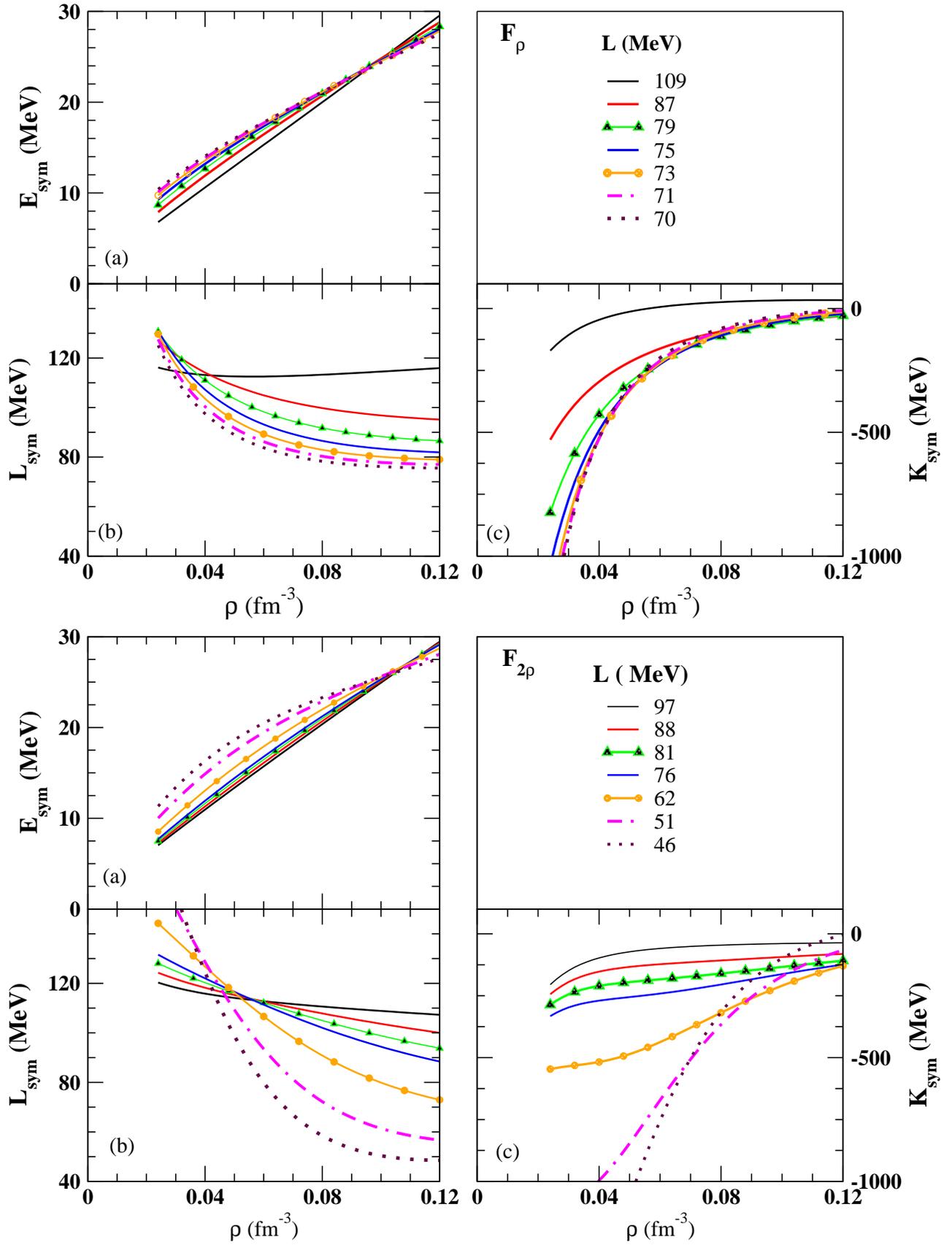

\begin{tabular}{cc}
\includegraphics[width=0.95\linewidth,angle=0]{fig2a.eps} \\
\includegraphics[width=0.95\linewidth,angle=0]{fig2b.eps} 
\end{tabular}
\caption{(Color online) (a) Symmetry energy  $E_{\rm sym}$, (b)  its slope  $L$ and  (c) curvature
  $K_{\rm sym}$, at subsaturation densities for the families $F_\rho$ (upper
  panels) and $F_{2\rho}$ (lower panels). }
\label{Fig:esym}
\end{figure*}

\begin{figure}
\includegraphics[width=1.1\linewidth,angle=0]{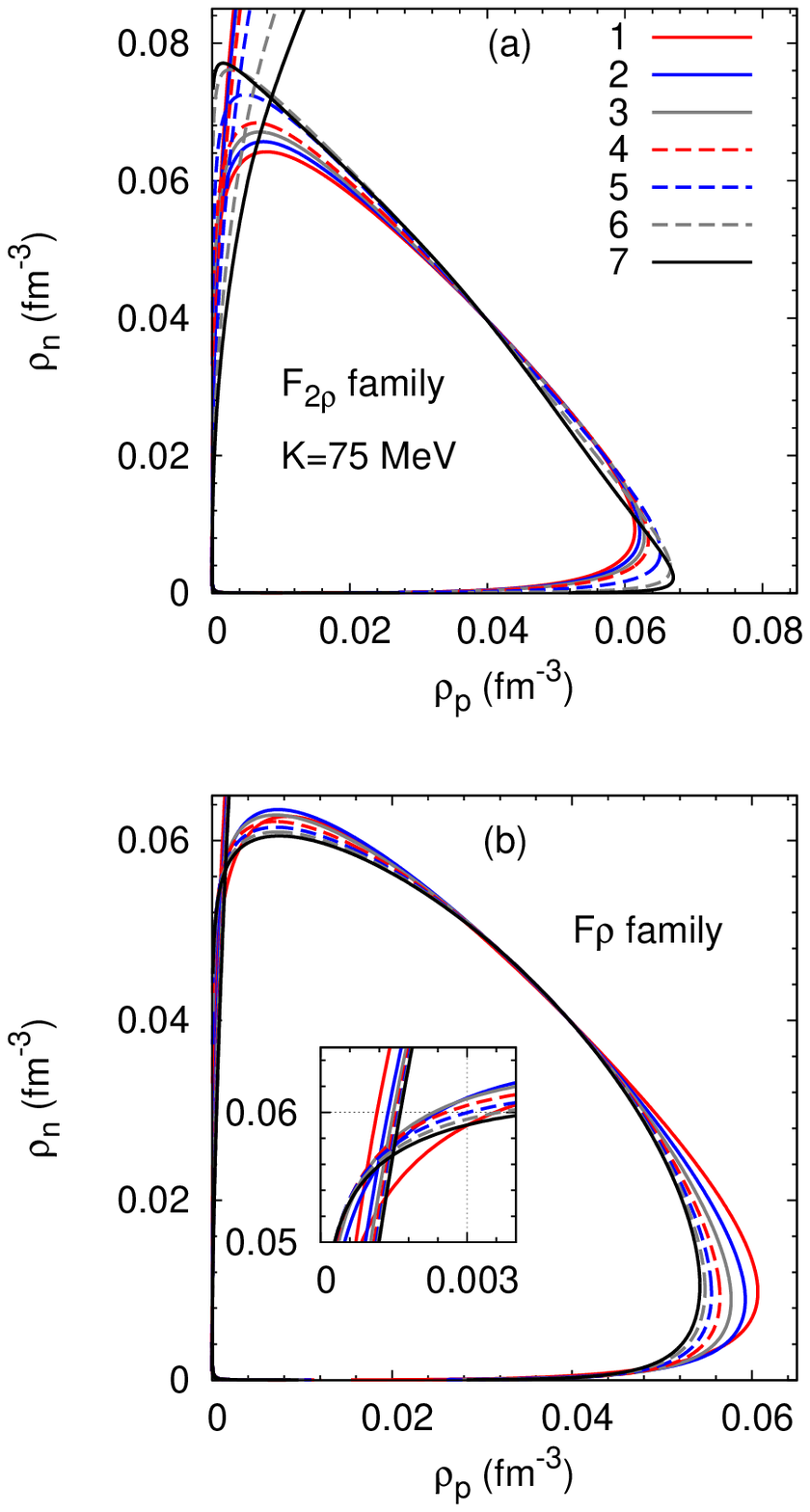} 
\caption{(Color online) Spinodal sections for the (a) $F_{2\rho}$ and (b) $F_{\rho}$ families. The EoS of
$\beta$-equilibrium stellar matter are also represented.}
\label{Fig:spin}
\end{figure}

In Fig. \ref{Fig:esym}, we plot for both families the symmetry energy
and its derivatives with respect to the density as a function of the
density at sub-saturation. It is clearly seen that the constraint
imposed on the neutron skin leads to the crossing of the curves
obtained with the  different
parametrizations of family $F_\rho$ for the
  symmetry energy and its slope, respectively,  below $0.1$fm$^{-3}$
and below $0.04$fm$^{-3}$, while for family $F_{2\rho}$ these
crossings occur above those densities. In particular, it is seen that
$K_{sym}$ is changing much slower at low densities within family
$F_{2\rho}$. These properties give rise to a different behavior of the
spinodals, see Fig. \ref{Fig:spin}. While for the family $F_{2\rho}$, the
spinodal regions are larger, the smaller the slope, $L$, a behavior
discussed in \cite{Ducoin11}, the contrary
occurs with family $F_\rho$. The effect of $L$ on the EoS of
$\beta$-equilibrium matter is the same: the smaller $L$, the larger the
proton fraction for a given density. These two effects add up for the
family $F_{2\rho}$, and  favor
a larger transition density, while for the family $F_\rho$, the two
effects act in opposite directions, and, as a result, the transition density does not change much. Let us point out that the behavior of the scalar density is purely
relativistic, and, therefore, Skyrme force behave all as  the $F_{2\rho}$ family, and not as the $F_{\rho}$ one. Most of the RMF models
describe their isovector channel through the $\rho$ meson alone, that brings
a baryonic density dependence on the energy and pressure, similar to
the one found with Skyrme interactions.

\begin{figure}
\begin{tabular}{c}
\includegraphics[width=0.5\textwidth]{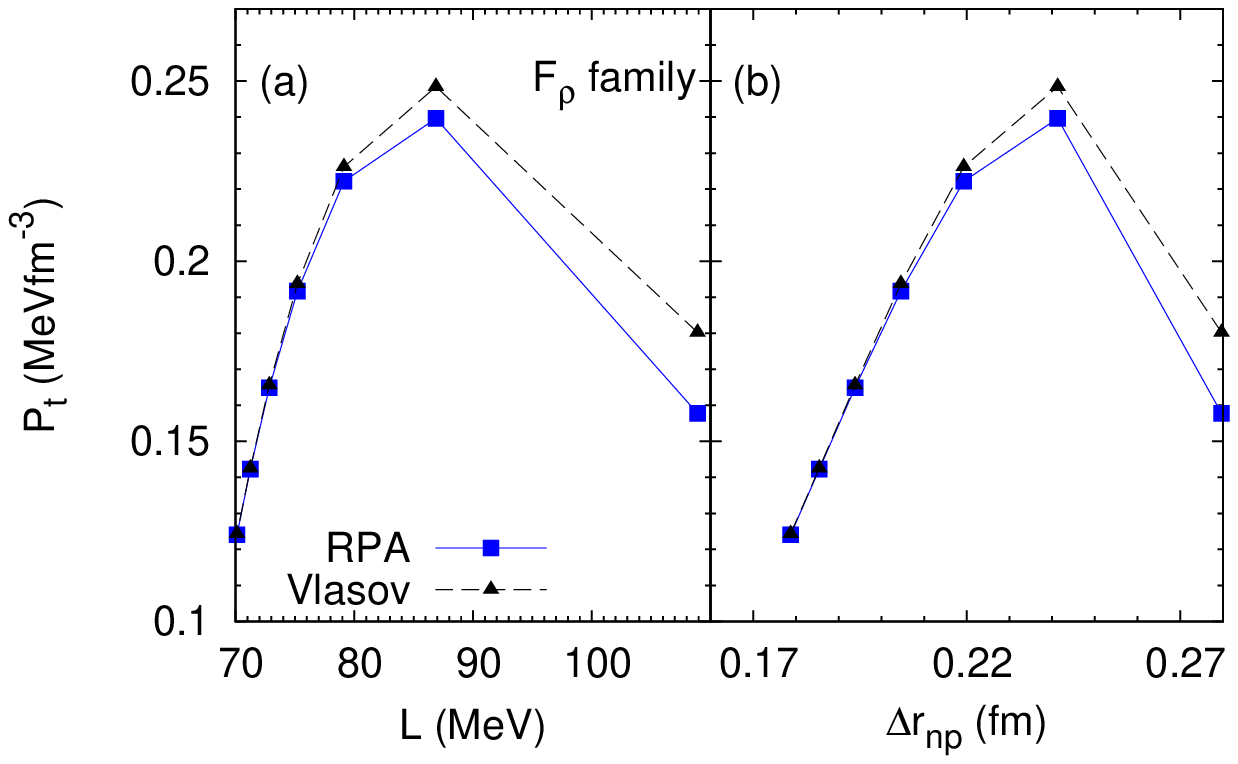}  \\
\includegraphics[width=0.5\textwidth]{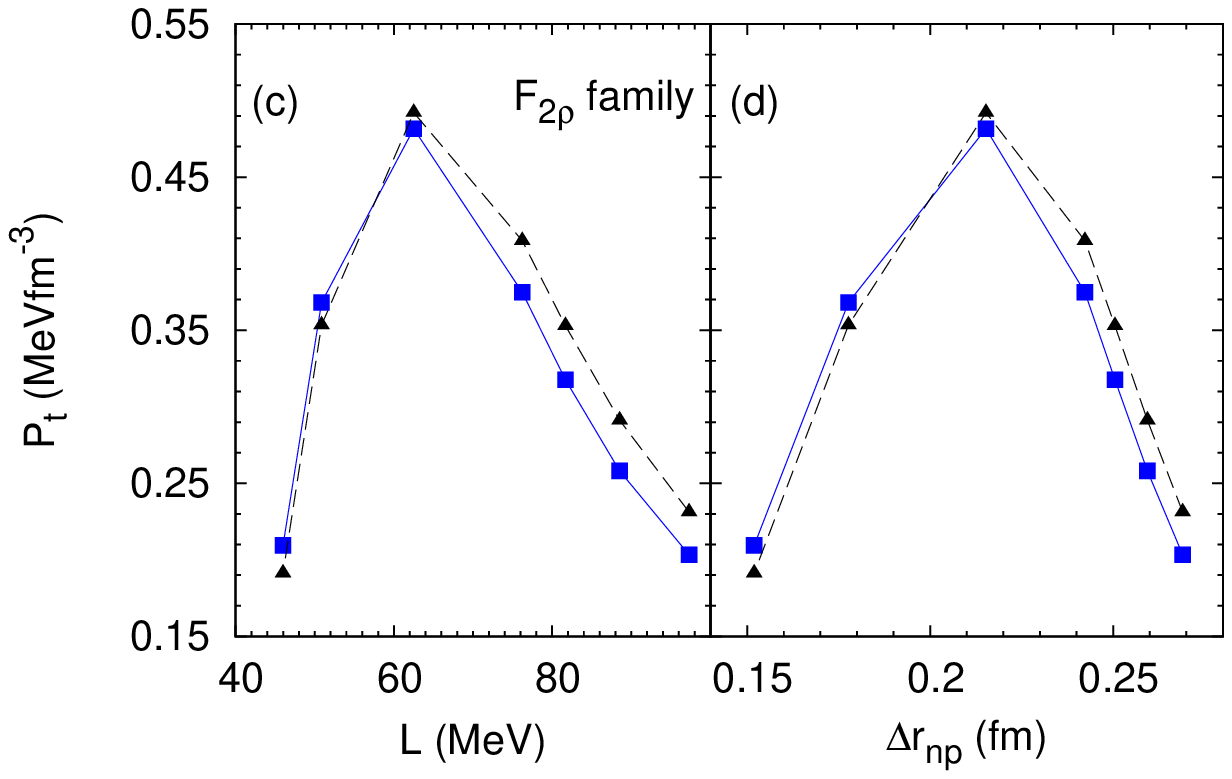}  
 \end{tabular}
\caption{(Color online) Crust-core transition pressures, $P_t$, for
  the F$_\rho$ family ((a) and (b)) and F$_{2\rho}$  family ((c) and (d)) as a
  function of $L$ (left panels) and  the neutron skin thickness,
  $\Delta r_{\rm np}$ (right panels) from RRPA, and Vlasov calculations.}
\label{Fig:PtMdlCom}
\end{figure}

In Fig.~\ref{Fig:PtMdlCom}, we show the transition pressure, $P_t$, as
a function of the slope of the symmetry energy, $L$, (left panels),
and as a function of the neutron skin thickness, $\Delta r_{\rm np}$, (right panels), for both families considered in this study.  In general, the predicted
$P_t$  depends significantly, not only on the isovector, but also on
the isoscalar mixing nonlinear terms used in the model. 
  As discussed previously
  in \cite{Ducoin-08,Ducoin11}, the behaviour of $P_t$ is not monotonic
  with $L$. For lower $L$ values, the pressure increases with
  $L$, followed by a steep decrease, for the larger $L$ values. For the
  family $F_\rho$, the pressure increases steadily, almost linearly,
  until $L=85$ MeV. This behaviour is similar to the one discussed in
  \cite{Moustakidis-10}.

In Ref. \cite{Piekarewicz-14}, it has been proposed that even with
entrainment, the crust is large enough to explain glitches,
choosing a parametrization that predicts a large pressure
at the crust-core transition. A density dependence of
the symmetry energy similar to the one of the family $F_{2\rho}$ was used, and transition pressures as large as $0.425$ and $0.550$ MeV/fm$^3$
for $L\sim 60$ MeV and $\Delta r_{\rm np}\sim 0.20-0.22$ fm, in accordance
with the results of family $F_{2\rho}$, were obtained. If, however, the density
dependence of the symmetry energy goes as the family $F_\rho$, a smaller
maximum pressure, $\sim 0.25$ MeV/fm$^3$, for $L\sim 85$ MeV  and
$\Delta r_{\rm np}\sim 0.24$ fm, would have been possible. In this
case, the crust would have not been enough to explain the glitches.

\begin{figure}
\epsfig{figure=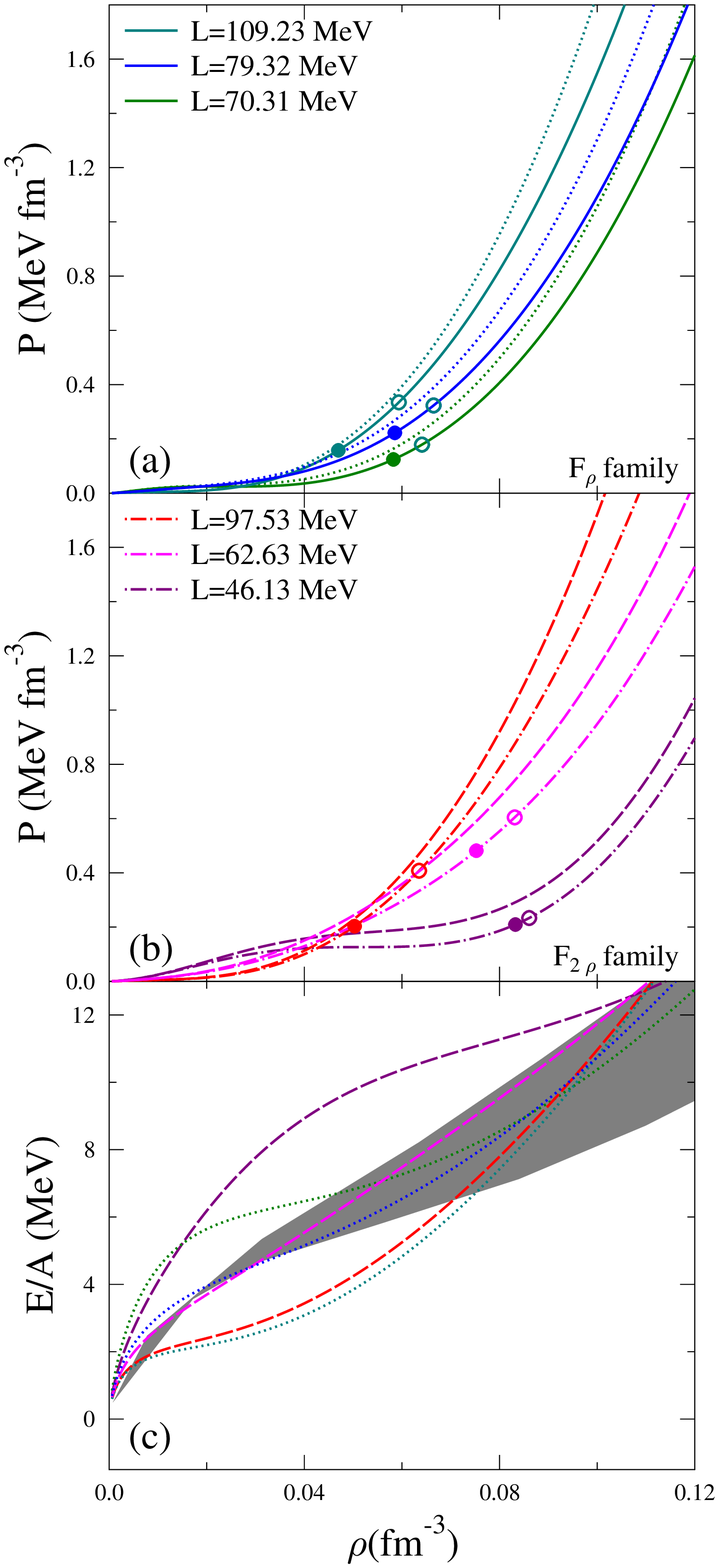,width=0.44\textwidth }
\caption{(Color online)  EOS for the (a) $F_\rho$  and (b) $F_{2\rho}$ 
families. The solid (dash-dotted) and  dotted (dashed) lines
are the EOSs for the $F_\rho$  ($F_{2\rho}$) families, corresponding to NS matter
and pure neutron matter, respectively.   Circles (full circles) in (a) and (b) correspond to
the transition pressure, $P_t$, calculated with the RRPA method for the case without (with) the
Coulomb contribution.  The energy per nucleon of the pure
neutron matter (c) for both families  is also displayed. The grey region shown is the pure
neutron matter result from Ref.~\cite{KTHS2013}.} 
\label{Fig:EOSLd}
\end{figure}

For completeness, we plot in Fig.~\ref{Fig:EOSLd} the EoS (upper panel) for neutron star matter (lines) and
pure neutron matter (dotted lines) at low densities and their
corresponding binding energies ($E/A$) (lower panel), for a few
$F_\rho$ and $F_{2 \rho}$ parameter sets. The  grey region in the
lower panel is the pure neutron matter result calculated from a chiral
effective field theory (see Ref.~\cite{KTHS2013} for details). The
full and empty circles in the upper panel correspond to the $P_t$
calculated from RRPA  with and without Coulomb contribution,
respectively. It can be seen in the upper panel of
Fig.~\ref{Fig:EOSLd} that the difference in $P_t$ appears more
significantly for relative  smaller $L$. The EoSs and the
corresponding $E/A$ for NS and pure neutron matter at low densities
vary significantly, depending on the parameter sets used. In general,
the binding energies for pure neutron matter calculated with $F_\rho$ and
$F_{2 \rho}$ families, with $L$ around 79.32 and  62.63  MeV,
respectively, are quite compatible with the calculations from chiral
effective field theory. 
However,
the significant difference of $P_t$ predicted by  $F_\rho$ and $F_{2\rho}$ families around this range of $L$ is due to the role of the  meson
mixed and self-interaction nonlinear terms in the EoS.

One consequence of the  different behavior of both families
 could be that, contrary to the proposed in \cite{Piekarewicz-14}, the
 crust might not be enough to describe glitches due to entrainment.
For the two parametrizations compatible with the
 chiral effective theories, we have
 obtained 0.22 and
0.49 MeV/fm$^{3}$ for the transition pressure  for
$F_\rho(L=79)$ and $F_{2\rho}(L=62)$, respectively.
Considering for  the moment of inertia
of the  crust, $I_{cr}$, the expression given in
\cite{Fattoyev-10a,Piekarewicz-14}, with $I_{cr}$  proportional to
$R^6_t \, P_t/R$, we get for a 1.4$M_\odot$ star,
$I_{cr}[F_\rho(L=79)]/I_{cr}[F_{2\rho}(L=62)]\sim 2/3$, indicating that the
contribution of the crust moment of inertia may be smaller than the
prediction of \cite{Piekarewicz-14}. The radius at the crust-core
transition was estimated taking the BPS EoS \cite{bps} for the outer
crust,
the FSU inner crust  obtained in the framework of a Thomas-Fermi calculation \cite{Grill-14} ,
 and
matching the homogeneous matter EoS to the inner crust EoS at the
crust-core transition.  We expect the inner crust of the FSU to be a good
choice because this model has a slope $L$ similar to the two models.
We
have obtained 
$R_t\left[F_\rho(L=79)\right]/R_t\left[F_{2\rho}(L=62)\right]\sim
1.03$.

\section{CONCLUSIONS}
\label{sec_conclu}

In the present work, the influence of  the density dependence of the
symmetry energy on properties such as the density and pressure at the
crust-core transition was analysed within two families of RMF models.
We have used three different methods to calculate the transition density:
the thermodynamical spinodal, the dynamical spinodal, within the Vlasov
formalism, and the RRPA. 
It was shown that the
last two methods give similar results, confirming previous studies
\cite{Ducoin-08a,Horowitz-08}. The thermodynamical spinodal is also
giving a good estimate of the transition density as already shown in
\cite{Ducoin11}, and involves simpler calculations.
The mixed terms involving the $\rho$-meson
and the $\sigma$ or the $\omega$ mesons allow the modification of the
density dependence of the symmetry energy.  The two families of the RMF
models differ in mixed non-linear meson terms in the isovector part of
the  Lagrangian density.  Both families of the RMF models  have
the same isoscalar properties, but the isovector channel is modified
through a $\sigma\rho^2$ term in the $F_\rho$ family, and a $\omega^2\rho^2$
term  in the $F_{2\rho}$ family. The parameters that describe the isovector
channel were appropriately adjusted,  so that different neutron
skin thicknesses were obtained for the $^{208}$Pb nucleus.  Since the
$\sigma$-meson field is proportional to the scalar density and the
$\omega$-meson to the baryonic density,  different behaviours of the crust-core transition properties were observed.

The scalar density is a relativistic quantity, and by performing an
expansion of the RMF energy density in powers of the Fermi momentum, it
has been shown in \cite{sw} that relativistic corrections coming from
the  Lorentz contraction factor in the scalar density have an effect
equivalent to repulsive many-body forces.  Due to  the much faster
increase of the scalar density at low densities, followed by a smoothing at larger, but still subsaturation densities,
the same neutron skin thicknesses were obtained for the $F_\rho$ family,
with larger values of  the slope $L$. Also the crust-core transition
was affected. The values of the pressure at the transition are lower,
there is no clear  decrease of the transition density with $L$, and the
transition pressure increases with the slope, $L$, for $L<85$ MeV. The
family $F_{2\rho}$, on the other hand, behaves as discussed in previous
works \cite{Vidana09,Ducoin-08,Ducoin11}, where both non-relativistic
and relativistic models have been considered, giving rise to similar
conclusions. The behavior of the family $F_{2\rho}$ is defined by the baryonic
density and, therefore, does not contain explicit relativistic effects.
If the density dependence of the symmetry energy should be defined by
the scalar density, we may expect smaller pressures at
the crust core transition. In this case, the crust would probably not be enough to describe glitches if entrainment is taken into account.

\section*{ACKNOWLEDGMENTS}

H.P. is supported by FCT under Project No. SFRH/BPD/95566/2013. Partial support comes from ``NewCompStar'', COST Action MP1304. A.S. is partly supported by the Research-Cluster-Grant-Program  of the University of Indonesia, under contract No.~1862/UN.R12/HKP.05.00/2015.


%


\begin{thebibliography}{50}%
\makeatletter
\providecommand \@ifxundefined [1]{%
 \@ifx{#1\undefined}
}%
\providecommand \@ifnum [1]{%
 \ifnum #1\expandafter \@firstoftwo
 \else \expandafter \@secondoftwo
 \fi
}%
\providecommand \@ifx [1]{%
 \ifx #1\expandafter \@firstoftwo
 \else \expandafter \@secondoftwo
 \fi
}%
\providecommand \natexlab [1]{#1}%
\providecommand \enquote  [1]{``#1''}%
\providecommand \bibnamefont  [1]{#1}%
\providecommand \bibfnamefont [1]{#1}%
\providecommand \citenamefont [1]{#1}%
\providecommand \href@noop [0]{\@secondoftwo}%
\providecommand \href [0]{\begingroup \@sanitize@url \@href}%
\providecommand \@href[1]{\@@startlink{#1}\@@href}%
\providecommand \@@href[1]{\endgroup#1\@@endlink}%
\providecommand \@sanitize@url [0]{\catcode `\\12\catcode `\$12\catcode
  `\&12\catcode `\#12\catcode `\^12\catcode `\_12\catcode `\%12\relax}%
\providecommand \@@startlink[1]{}%
\providecommand \@@endlink[0]{}%
\providecommand \url  [0]{\begingroup\@sanitize@url \@url }%
\providecommand \@url [1]{\endgroup\@href {#1}{\urlprefix }}%
\providecommand \urlprefix  [0]{URL }%
\providecommand \Eprint [0]{\href }%
\providecommand \doibase [0]{http://dx.doi.org/}%
\providecommand \selectlanguage [0]{\@gobble}%
\providecommand \bibinfo  [0]{\@secondoftwo}%
\providecommand \bibfield  [0]{\@secondoftwo}%
\providecommand \translation [1]{[#1]}%
\providecommand \BibitemOpen [0]{}%
\providecommand \bibitemStop [0]{}%
\providecommand \bibitemNoStop [0]{.\EOS\space}%
\providecommand \EOS [0]{\spacefactor3000\relax}%
\providecommand \BibitemShut  [1]{\csname bibitem#1\endcsname}%
\let\auto@bib@innerbib\@empty
\bibitem [{\citenamefont {Steiner}\ \emph {et~al.}(2005)\citenamefont
  {Steiner}, \citenamefont {Prakash}, \citenamefont {Lattimer},\ and\
  \citenamefont {Ellis}}]{Steiner-05}%
  \BibitemOpen
  \bibfield  {author} {\bibinfo {author} {\bibfnamefont {A.~W.}\ \bibnamefont
  {Steiner}}, \bibinfo {author} {\bibfnamefont {M.}~\bibnamefont {Prakash}},
  \bibinfo {author} {\bibfnamefont {J.~M.}\ \bibnamefont {Lattimer}}, \ and\
  \bibinfo {author} {\bibfnamefont {P.~J.}\ \bibnamefont {Ellis}},\ }\href@noop
  {} {\bibfield  {journal} {\bibinfo  {journal} {Phys.\ Rep.}\ }\textbf
  {\bibinfo {volume} {411}},\ \bibinfo {pages} {325} (\bibinfo {year}
  {2005})}\BibitemShut {NoStop}%
\bibitem [{\citenamefont {Li}\ \emph {et~al.}(2008)\citenamefont {Li},
  \citenamefont {Chen},\ and\ \citenamefont {Ko}}]{Li08}%
  \BibitemOpen
  \bibfield  {author} {\bibinfo {author} {\bibfnamefont {B.-A.}\ \bibnamefont
  {Li}}, \bibinfo {author} {\bibfnamefont {L.-W.}\ \bibnamefont {Chen}}, \ and\
  \bibinfo {author} {\bibfnamefont {C.-M.}\ \bibnamefont {Ko}},\ }\href@noop {}
  {\bibfield  {journal} {\bibinfo  {journal} {Phys.\ Rep.}\ }\textbf {\bibinfo
  {volume} {464}},\ \bibinfo {pages} {113} (\bibinfo {year}
  {2008})}\BibitemShut {NoStop}%
\bibitem [{\citenamefont {Tsang}\ \emph {et~al.}(2012)\citenamefont {Tsang},
  \citenamefont {Stone}, \citenamefont {Camera}, \citenamefont {Danielewicz},
  \citenamefont {Gandolfi}, \citenamefont {Hebeler}, \citenamefont {Horowitz},
  \citenamefont {Lee}, \citenamefont {Lynch}, \citenamefont {Kohley},
  \citenamefont {Lemmon}, \citenamefont {M{\"o}ller}, \citenamefont {Murakami},
  \citenamefont {Riordan}, \citenamefont {Roca-Maza}, \citenamefont
  {Sammarruca}, \citenamefont {Steiner}, \citenamefont {Vida{\~n}a},\ and\
  \citenamefont {Yennello}}]{Tsang-12}%
  \BibitemOpen
  \bibfield  {author} {\bibinfo {author} {\bibfnamefont {M.~B.}\ \bibnamefont
  {Tsang}}, \bibinfo {author} {\bibfnamefont {J.~R.}\ \bibnamefont {Stone}},
  \bibinfo {author} {\bibfnamefont {F.}~\bibnamefont {Camera}}, \bibinfo
  {author} {\bibfnamefont {P.}~\bibnamefont {Danielewicz}}, \bibinfo {author}
  {\bibfnamefont {S.}~\bibnamefont {Gandolfi}}, \bibinfo {author}
  {\bibfnamefont {K.}~\bibnamefont {Hebeler}}, \bibinfo {author} {\bibfnamefont
  {C.~J.}\ \bibnamefont {Horowitz}}, \bibinfo {author} {\bibfnamefont
  {J.}~\bibnamefont {Lee}}, \bibinfo {author} {\bibfnamefont {W.~G.}\
  \bibnamefont {Lynch}}, \bibinfo {author} {\bibfnamefont {Z.}~\bibnamefont
  {Kohley}}, \bibinfo {author} {\bibfnamefont {R.}~\bibnamefont {Lemmon}},
  \bibinfo {author} {\bibfnamefont {P.}~\bibnamefont {M{\"o}ller}}, \bibinfo
  {author} {\bibfnamefont {T.}~\bibnamefont {Murakami}}, \bibinfo {author}
  {\bibfnamefont {S.}~\bibnamefont {Riordan}}, \bibinfo {author} {\bibfnamefont
  {X.}~\bibnamefont {Roca-Maza}}, \bibinfo {author} {\bibfnamefont
  {F.}~\bibnamefont {Sammarruca}}, \bibinfo {author} {\bibfnamefont {A.~W.}\
  \bibnamefont {Steiner}}, \bibinfo {author} {\bibfnamefont {I.}~\bibnamefont
  {Vida{\~n}a}}, \ and\ \bibinfo {author} {\bibfnamefont {S.~J.}\ \bibnamefont
  {Yennello}},\ }\href@noop {} {\bibfield  {journal} {\bibinfo  {journal}
  {Phys.\ Rev.\ C}\ }\textbf {\bibinfo {volume} {86}},\ \bibinfo {pages}
  {015803} (\bibinfo {year} {2012})}\BibitemShut {NoStop}%
\bibitem [{\citenamefont {Garg}\ \emph {et~al.}(2007)\citenamefont {Garg},
  \citenamefont {Li}, \citenamefont {Okumura}, \citenamefont {Akimune},
  \citenamefont {Fujiwara}, \citenamefont {Harakeh}, \citenamefont {Hashimoto},
  \citenamefont {Itoh}, \citenamefont {Iwao}, \citenamefont {Kawabata},
  \citenamefont {Kawase}, \citenamefont {Liu}, \citenamefont {Marks},
  \citenamefont {Murakami}, \citenamefont {Nakanishi}, \citenamefont {Nayak},
  \citenamefont {Rao}, \citenamefont {Sakaguchi}, \citenamefont {Terashima},
  \citenamefont {Uchida}, \citenamefont {Yasuda}, \citenamefont {Yosoi},\ and\
  \citenamefont {Zenihirof}}]{Garg-07}%
  \BibitemOpen
  \bibfield  {author} {\bibinfo {author} {\bibfnamefont {U.}~\bibnamefont
  {Garg}}, \bibinfo {author} {\bibfnamefont {T.}~\bibnamefont {Li}}, \bibinfo
  {author} {\bibfnamefont {S.}~\bibnamefont {Okumura}}, \bibinfo {author}
  {\bibfnamefont {H.}~\bibnamefont {Akimune}}, \bibinfo {author} {\bibfnamefont
  {M.}~\bibnamefont {Fujiwara}}, \bibinfo {author} {\bibfnamefont
  {M.}~\bibnamefont {Harakeh}}, \bibinfo {author} {\bibfnamefont
  {H.}~\bibnamefont {Hashimoto}}, \bibinfo {author} {\bibfnamefont
  {M.}~\bibnamefont {Itoh}}, \bibinfo {author} {\bibfnamefont {Y.}~\bibnamefont
  {Iwao}}, \bibinfo {author} {\bibfnamefont {T.}~\bibnamefont {Kawabata}},
  \bibinfo {author} {\bibfnamefont {K.}~\bibnamefont {Kawase}}, \bibinfo
  {author} {\bibfnamefont {Y.}~\bibnamefont {Liu}}, \bibinfo {author}
  {\bibfnamefont {R.}~\bibnamefont {Marks}}, \bibinfo {author} {\bibfnamefont
  {T.}~\bibnamefont {Murakami}}, \bibinfo {author} {\bibfnamefont
  {K.}~\bibnamefont {Nakanishi}}, \bibinfo {author} {\bibfnamefont
  {B.}~\bibnamefont {Nayak}}, \bibinfo {author} {\bibfnamefont {P.~M.}\
  \bibnamefont {Rao}}, \bibinfo {author} {\bibfnamefont {H.}~\bibnamefont
  {Sakaguchi}}, \bibinfo {author} {\bibfnamefont {Y.}~\bibnamefont
  {Terashima}}, \bibinfo {author} {\bibfnamefont {M.}~\bibnamefont {Uchida}},
  \bibinfo {author} {\bibfnamefont {Y.}~\bibnamefont {Yasuda}}, \bibinfo
  {author} {\bibfnamefont {M.}~\bibnamefont {Yosoi}}, \ and\ \bibinfo {author}
  {\bibfnamefont {J.}~\bibnamefont {Zenihirof}},\ }\href@noop {} {\bibfield
  {journal} {\bibinfo  {journal} {Nucl.\ Phys.\ A}\ }\textbf {\bibinfo {volume}
  {788}},\ \bibinfo {pages} {36} (\bibinfo {year} {2007})}\BibitemShut
  {NoStop}%
\bibitem [{\citenamefont {Danielewicz}\ and\ \citenamefont
  {Lee}(2009)}]{Danielewicz09}%
  \BibitemOpen
  \bibfield  {author} {\bibinfo {author} {\bibfnamefont {P.}~\bibnamefont
  {Danielewicz}}\ and\ \bibinfo {author} {\bibfnamefont {J.}~\bibnamefont
  {Lee}},\ }\href@noop {} {\bibfield  {journal} {\bibinfo  {journal} {Nucl.\
  Phys.\ A}\ }\textbf {\bibinfo {volume} {818}},\ \bibinfo {pages} {36}
  (\bibinfo {year} {2009})}\BibitemShut {NoStop}%
\bibitem [{\citenamefont {Li}\ \emph {et~al.}(2005)\citenamefont {Li},
  \citenamefont {Yong},\ and\ \citenamefont {Zuo}}]{Li05}%
  \BibitemOpen
  \bibfield  {author} {\bibinfo {author} {\bibfnamefont {B.-A.}\ \bibnamefont
  {Li}}, \bibinfo {author} {\bibfnamefont {G.-C.}\ \bibnamefont {Yong}}, \ and\
  \bibinfo {author} {\bibfnamefont {W.}~\bibnamefont {Zuo}},\ }\href@noop {}
  {\bibfield  {journal} {\bibinfo  {journal} {Phys.\ Rev.\ C}\ }\textbf
  {\bibinfo {volume} {71}},\ \bibinfo {pages} {014608} (\bibinfo {year}
  {2005})}\BibitemShut {NoStop}%
\bibitem [{\citenamefont {Fuchs}(2006)}]{Fuchs06}%
  \BibitemOpen
  \bibfield  {author} {\bibinfo {author} {\bibfnamefont {C.}~\bibnamefont
  {Fuchs}},\ }\href@noop {} {\bibfield  {journal} {\bibinfo  {journal} {Prog.\
  Part.\ Nucl.\ Phys.}\ }\textbf {\bibinfo {volume} {56}},\ \bibinfo {pages}
  {1} (\bibinfo {year} {2006})}\BibitemShut {NoStop}%
\bibitem [{\citenamefont {Brown}(2000)}]{Brown00}%
  \BibitemOpen
  \bibfield  {author} {\bibinfo {author} {\bibfnamefont {B.~A.}\ \bibnamefont
  {Brown}},\ }\href@noop {} {\bibfield  {journal} {\bibinfo  {journal} {Phys.\
  Rev.\ Lett.}\ }\textbf {\bibinfo {volume} {85}},\ \bibinfo {pages} {5296}
  (\bibinfo {year} {2000})}\BibitemShut {NoStop}%
\bibitem [{\citenamefont {Typel}\ and\ \citenamefont {Brown}(2001)}]{Typel01}%
  \BibitemOpen
  \bibfield  {author} {\bibinfo {author} {\bibfnamefont {S.}~\bibnamefont
  {Typel}}\ and\ \bibinfo {author} {\bibfnamefont {B.~A.}\ \bibnamefont
  {Brown}},\ }\href@noop {} {\bibfield  {journal} {\bibinfo  {journal} {Phys.\
  Rev.\ C}\ }\textbf {\bibinfo {volume} {64}},\ \bibinfo {pages} {027302}
  (\bibinfo {year} {2001})}\BibitemShut {NoStop}%
\bibitem [{\citenamefont {Horowitz}\ and\ \citenamefont
  {Piekarewicz}(2001)}]{Horowitz01}%
  \BibitemOpen
  \bibfield  {author} {\bibinfo {author} {\bibfnamefont {C.~J.}\ \bibnamefont
  {Horowitz}}\ and\ \bibinfo {author} {\bibfnamefont {J.}~\bibnamefont
  {Piekarewicz}},\ }\href@noop {} {\bibfield  {journal} {\bibinfo  {journal}
  {Phys.\ Rev.\ Lett.}\ }\textbf {\bibinfo {volume} {86}},\ \bibinfo {pages}
  {5647} (\bibinfo {year} {2001})}\BibitemShut {NoStop}%
\bibitem [{\citenamefont {Vida{\~n}a}\ \emph {et~al.}(2009)\citenamefont
  {Vida{\~n}a}, \citenamefont {Provid\^encia}, \citenamefont {Polls},\ and\
  \citenamefont {Rios}}]{Vidana09}%
  \BibitemOpen
  \bibfield  {author} {\bibinfo {author} {\bibfnamefont {I.}~\bibnamefont
  {Vida{\~n}a}}, \bibinfo {author} {\bibfnamefont {C.}~\bibnamefont
  {Provid\^encia}}, \bibinfo {author} {\bibfnamefont {A.}~\bibnamefont
  {Polls}}, \ and\ \bibinfo {author} {\bibfnamefont {A.}~\bibnamefont {Rios}},\
  }\href@noop {} {\bibfield  {journal} {\bibinfo  {journal} {Phys.\ Rev.\ C}\
  }\textbf {\bibinfo {volume} {80}},\ \bibinfo {pages} {045806} (\bibinfo
  {year} {2009})}\BibitemShut {NoStop}%
\bibitem [{\citenamefont {Link}\ \emph {et~al.}(1999)\citenamefont {Link},
  \citenamefont {Epstein},\ and\ \citenamefont {Lattimer}}]{Link-99}%
  \BibitemOpen
  \bibfield  {author} {\bibinfo {author} {\bibfnamefont {B.}~\bibnamefont
  {Link}}, \bibinfo {author} {\bibfnamefont {R.~I.}\ \bibnamefont {Epstein}}, \
  and\ \bibinfo {author} {\bibfnamefont {J.~M.}\ \bibnamefont {Lattimer}},\
  }\href@noop {} {\bibfield  {journal} {\bibinfo  {journal} {Phys.\ Rev.\
  Lett.}\ }\textbf {\bibinfo {volume} {83}},\ \bibinfo {pages} {3362} (\bibinfo
  {year} {1999})}\BibitemShut {NoStop}%
\bibitem [{\citenamefont {Chamel}(2013)}]{Chamel-13}%
  \BibitemOpen
  \bibfield  {author} {\bibinfo {author} {\bibfnamefont {N.}~\bibnamefont
  {Chamel}},\ }\href@noop {} {\bibfield  {journal} {\bibinfo  {journal} {Phys.\
  Rev.\ Lett.}\ }\textbf {\bibinfo {volume} {110}},\ \bibinfo {pages} {011101}
  (\bibinfo {year} {2013})}\BibitemShut {NoStop}%
\bibitem [{\citenamefont {Andersson}\ \emph {et~al.}(2012)\citenamefont
  {Andersson}, \citenamefont {Glampedakis}, \citenamefont {Ho},\ and\
  \citenamefont {Espinoza}}]{Andersson-12}%
  \BibitemOpen
  \bibfield  {author} {\bibinfo {author} {\bibfnamefont {N.}~\bibnamefont
  {Andersson}}, \bibinfo {author} {\bibfnamefont {K.}~\bibnamefont
  {Glampedakis}}, \bibinfo {author} {\bibfnamefont {W.}~\bibnamefont {Ho}}, \
  and\ \bibinfo {author} {\bibfnamefont {C.}~\bibnamefont {Espinoza}},\
  }\href@noop {} {\bibfield  {journal} {\bibinfo  {journal} {Phys.\ Rev.\
  Lett.}\ }\textbf {\bibinfo {volume} {109}},\ \bibinfo {pages} {241103}
  (\bibinfo {year} {2012})}\BibitemShut {NoStop}%
\bibitem [{\citenamefont {Piekarewicz}\ \emph {et~al.}(2014)\citenamefont
  {Piekarewicz}, \citenamefont {Fattoyev},\ and\ \citenamefont
  {Horowitz}}]{Piekarewicz-14}%
  \BibitemOpen
  \bibfield  {author} {\bibinfo {author} {\bibfnamefont {J.}~\bibnamefont
  {Piekarewicz}}, \bibinfo {author} {\bibfnamefont {F.~J.}\ \bibnamefont
  {Fattoyev}}, \ and\ \bibinfo {author} {\bibfnamefont {C.~J.}\ \bibnamefont
  {Horowitz}},\ }\href@noop {} {\bibfield  {journal} {\bibinfo  {journal}
  {Phys.\ Rev.\ C}\ }\textbf {\bibinfo {volume} {90}},\ \bibinfo {pages}
  {015803} (\bibinfo {year} {2014})}\BibitemShut {NoStop}%
\bibitem [{\citenamefont {M{\"u}ller}\ and\ \citenamefont
  {Serot}(1996)}]{Mueller-96}%
  \BibitemOpen
  \bibfield  {author} {\bibinfo {author} {\bibfnamefont {H.}~\bibnamefont
  {M{\"u}ller}}\ and\ \bibinfo {author} {\bibfnamefont {B.~D.}\ \bibnamefont
  {Serot}},\ }\href@noop {} {\bibfield  {journal} {\bibinfo  {journal} {Nuc.\
  Phys.\ A}\ }\textbf {\bibinfo {volume} {606}},\ \bibinfo {pages} {508}
  (\bibinfo {year} {1996})}\BibitemShut {NoStop}%
\bibitem [{\citenamefont {Typel}\ and\ \citenamefont
  {Wolter}(1999)}]{Typel-99}%
  \BibitemOpen
  \bibfield  {author} {\bibinfo {author} {\bibfnamefont {S.}~\bibnamefont
  {Typel}}\ and\ \bibinfo {author} {\bibfnamefont {H.~H.}\ \bibnamefont
  {Wolter}},\ }\href@noop {} {\bibfield  {journal} {\bibinfo  {journal} {Nucl.\
  Phys.\ A}\ }\textbf {\bibinfo {volume} {656}},\ \bibinfo {pages} {331}
  (\bibinfo {year} {1999})}\BibitemShut {NoStop}%
\bibitem [{\citenamefont {{Agrawal}}(2010)}]{Agrawal-10}%
  \BibitemOpen
  \bibfield  {author} {\bibinfo {author} {\bibfnamefont {B.~K.}\ \bibnamefont
  {{Agrawal}}},\ }\href {\doibase 10.1103/PhysRevC.81.034323} {\bibfield
  {journal} {\bibinfo  {journal} {\prc}\ }\textbf {\bibinfo {volume} {81}},\
  \bibinfo {eid} {034323} (\bibinfo {year} {2010})}\BibitemShut {NoStop}%
\bibitem [{\citenamefont {Serot}\ and\ \citenamefont {Walecka}(1997)}]{sw}%
  \BibitemOpen
  \bibfield  {author} {\bibinfo {author} {\bibfnamefont {B.~D.}\ \bibnamefont
  {Serot}}\ and\ \bibinfo {author} {\bibfnamefont {J.~D.}\ \bibnamefont
  {Walecka}},\ }\href@noop {} {\bibfield  {journal} {\bibinfo  {journal} {Int.\
  J.\ Mod.\ Phys.}\ }\textbf {\bibinfo {volume} {E6}},\ \bibinfo {pages} {515}
  (\bibinfo {year} {1997})}\BibitemShut {NoStop}%
\bibitem [{\citenamefont {Provid\^encia}\ \emph
  {et~al.}(2006{\natexlab{a}})\citenamefont {Provid\^encia}, \citenamefont
  {Brito}, \citenamefont {Santos}, \citenamefont {Menezes},\ and\ \citenamefont
  {Avancini}}]{ProvidenciaC-06a}%
  \BibitemOpen
  \bibfield  {author} {\bibinfo {author} {\bibfnamefont {C.}~\bibnamefont
  {Provid\^encia}}, \bibinfo {author} {\bibfnamefont {L.}~\bibnamefont
  {Brito}}, \bibinfo {author} {\bibfnamefont {A.~M.~S.}\ \bibnamefont
  {Santos}}, \bibinfo {author} {\bibfnamefont {D.~P.}\ \bibnamefont {Menezes}},
  \ and\ \bibinfo {author} {\bibfnamefont {S.~S.}\ \bibnamefont {Avancini}},\
  }\href@noop {} {\bibfield  {journal} {\bibinfo  {journal} {Phys.\ Rev.\ C}\
  }\textbf {\bibinfo {volume} {74}},\ \bibinfo {pages} {045802} (\bibinfo
  {year} {2006}{\natexlab{a}})}\BibitemShut {NoStop}%
\bibitem [{\citenamefont {Agrawal}\ \emph {et~al.}(2012)\citenamefont
  {Agrawal}, \citenamefont {Sulaksono},\ and\ \citenamefont
  {Reinhard}}]{ASR2012}%
  \BibitemOpen
  \bibfield  {author} {\bibinfo {author} {\bibfnamefont {B.~K.}\ \bibnamefont
  {Agrawal}}, \bibinfo {author} {\bibfnamefont {A.}~\bibnamefont {Sulaksono}},
  \ and\ \bibinfo {author} {\bibfnamefont {P.}~\bibnamefont {Reinhard}},\
  }\href@noop {} {\bibfield  {journal} {\bibinfo  {journal} {Nucl.\ Phys.\ A}\
  }\textbf {\bibinfo {volume} {882}},\ \bibinfo {pages} {1} (\bibinfo {year}
  {2012})}\BibitemShut {NoStop}%
\bibitem [{\citenamefont {Sulaksono}\ and\ \citenamefont {Mart}(2006)}]{AT06}%
  \BibitemOpen
  \bibfield  {author} {\bibinfo {author} {\bibfnamefont {A.}~\bibnamefont
  {Sulaksono}}\ and\ \bibinfo {author} {\bibfnamefont {T.}~\bibnamefont
  {Mart}},\ }\href@noop {} {\bibfield  {journal} {\bibinfo  {journal} {Phys.\
  Rev.\ C}\ }\textbf {\bibinfo {volume} {74}},\ \bibinfo {pages} {045806}
  (\bibinfo {year} {2006})}\BibitemShut {NoStop}%
\bibitem [{\citenamefont {Pais}\ \emph {et~al.}(2010)\citenamefont {Pais},
  \citenamefont {Santos}, \citenamefont {Brito},\ and\ \citenamefont
  {Provid\^encia}}]{Pais10}%
  \BibitemOpen
  \bibfield  {author} {\bibinfo {author} {\bibfnamefont {H.}~\bibnamefont
  {Pais}}, \bibinfo {author} {\bibfnamefont {A.}~\bibnamefont {Santos}},
  \bibinfo {author} {\bibfnamefont {L.}~\bibnamefont {Brito}}, \ and\ \bibinfo
  {author} {\bibfnamefont {C.}~\bibnamefont {Provid\^encia}},\ }\href@noop {}
  {\bibfield  {journal} {\bibinfo  {journal} {Phys.\ Rev.\ C}\ }\textbf
  {\bibinfo {volume} {82}},\ \bibinfo {pages} {025801} (\bibinfo {year}
  {2010})}\BibitemShut {NoStop}%
\bibitem [{\citenamefont {Avancini}\ \emph {et~al.}(2010)\citenamefont
  {Avancini}, \citenamefont {Chiacchiera}, \citenamefont {Menezes},\ and\
  \citenamefont {Provid\^encia}}]{Avancini-10}%
  \BibitemOpen
  \bibfield  {author} {\bibinfo {author} {\bibfnamefont {S.~S.}\ \bibnamefont
  {Avancini}}, \bibinfo {author} {\bibfnamefont {S.}~\bibnamefont
  {Chiacchiera}}, \bibinfo {author} {\bibfnamefont {D.~P.}\ \bibnamefont
  {Menezes}}, \ and\ \bibinfo {author} {\bibfnamefont {C.}~\bibnamefont
  {Provid\^encia}},\ }\href@noop {} {\bibfield  {journal} {\bibinfo  {journal}
  {Phys.\ Rev.\ C}\ }\textbf {\bibinfo {volume} {82}},\ \bibinfo {pages}
  {055807} (\bibinfo {year} {2010})}\BibitemShut {NoStop}%
\bibitem [{\citenamefont {Ducoin}\ \emph
  {et~al.}(2008{\natexlab{a}})\citenamefont {Ducoin}, \citenamefont
  {Provid\^encia}, \citenamefont {Santos}, \citenamefont {Brito},\ and\
  \citenamefont {Chomaz}}]{Ducoin-08}%
  \BibitemOpen
  \bibfield  {author} {\bibinfo {author} {\bibfnamefont {C.}~\bibnamefont
  {Ducoin}}, \bibinfo {author} {\bibfnamefont {C.}~\bibnamefont
  {Provid\^encia}}, \bibinfo {author} {\bibfnamefont {A.~M.}\ \bibnamefont
  {Santos}}, \bibinfo {author} {\bibfnamefont {L.}~\bibnamefont {Brito}}, \
  and\ \bibinfo {author} {\bibfnamefont {P.}~\bibnamefont {Chomaz}},\
  }\href@noop {} {\bibfield  {journal} {\bibinfo  {journal} {Phys.\ Rev.\ C}\
  }\textbf {\bibinfo {volume} {78}},\ \bibinfo {pages} {055801} (\bibinfo
  {year} {2008}{\natexlab{a}})}\BibitemShut {NoStop}%
\bibitem [{\citenamefont {Ducoin}\ \emph {et~al.}(2011)\citenamefont {Ducoin},
  \citenamefont {Margueron}, \citenamefont {Provid\^encia},\ and\ \citenamefont
  {Vida{\~n}a}}]{Ducoin11}%
  \BibitemOpen
  \bibfield  {author} {\bibinfo {author} {\bibfnamefont {C.}~\bibnamefont
  {Ducoin}}, \bibinfo {author} {\bibfnamefont {J.}~\bibnamefont {Margueron}},
  \bibinfo {author} {\bibfnamefont {C.}~\bibnamefont {Provid\^encia}}, \ and\
  \bibinfo {author} {\bibfnamefont {I.}~\bibnamefont {Vida{\~n}a}},\
  }\href@noop {} {\bibfield  {journal} {\bibinfo  {journal} {Phys.\ Rev.\ C}\
  }\textbf {\bibinfo {volume} {83}},\ \bibinfo {pages} {045810} (\bibinfo
  {year} {2011})}\BibitemShut {NoStop}%
\bibitem [{\citenamefont {Walecka}(1974)}]{Walecka-74}%
  \BibitemOpen
  \bibfield  {author} {\bibinfo {author} {\bibfnamefont {J.~D.}\ \bibnamefont
  {Walecka}},\ }\href@noop {} {\bibfield  {journal} {\bibinfo  {journal} {Ann.\
  Phys.}\ }\textbf {\bibinfo {volume} {83}},\ \bibinfo {pages} {491} (\bibinfo
  {year} {1974})}\BibitemShut {NoStop}%
\bibitem [{\citenamefont {{Furnstahl}}(2002)}]{Furnstahl-02}%
  \BibitemOpen
  \bibfield  {author} {\bibinfo {author} {\bibfnamefont {R.~J.}\ \bibnamefont
  {{Furnstahl}}},\ }\href {\doibase 10.1016/S0375-9474(02)00867-9} {\bibfield
  {journal} {\bibinfo  {journal} {Nuclear Physics A}\ }\textbf {\bibinfo
  {volume} {706}},\ \bibinfo {pages} {85} (\bibinfo {year} {2002})}\BibitemShut
  {NoStop}%
\bibitem [{\citenamefont {{Sil}}\ \emph {et~al.}(2005)\citenamefont {{Sil}},
  \citenamefont {{Centelles}}, \citenamefont {{Vi{\~n}as}},\ and\ \citenamefont
  {{Piekarewicz}}}]{Sil-05}%
  \BibitemOpen
  \bibfield  {author} {\bibinfo {author} {\bibfnamefont {T.}~\bibnamefont
  {{Sil}}}, \bibinfo {author} {\bibfnamefont {M.}~\bibnamefont {{Centelles}}},
  \bibinfo {author} {\bibfnamefont {X.}~\bibnamefont {{Vi{\~n}as}}}, \ and\
  \bibinfo {author} {\bibfnamefont {J.}~\bibnamefont {{Piekarewicz}}},\ }\href
  {\doibase 10.1103/PhysRevC.71.045502} {\bibfield  {journal} {\bibinfo
  {journal} {\prc}\ }\textbf {\bibinfo {volume} {71}},\ \bibinfo {eid} {045502}
  (\bibinfo {year} {2005})}\BibitemShut {NoStop}%
\bibitem [{\citenamefont {{Dhiman}}\ \emph {et~al.}(2007)\citenamefont
  {{Dhiman}}, \citenamefont {{Kumar}},\ and\ \citenamefont
  {{Agrawal}}}]{Dhiman-07}%
  \BibitemOpen
  \bibfield  {author} {\bibinfo {author} {\bibfnamefont {S.~K.}\ \bibnamefont
  {{Dhiman}}}, \bibinfo {author} {\bibfnamefont {R.}~\bibnamefont {{Kumar}}}, \
  and\ \bibinfo {author} {\bibfnamefont {B.~K.}\ \bibnamefont {{Agrawal}}},\
  }\href {\doibase 10.1103/PhysRevC.76.045801} {\bibfield  {journal} {\bibinfo
  {journal} {\prc}\ }\textbf {\bibinfo {volume} {76}},\ \bibinfo {eid} {045801}
  (\bibinfo {year} {2007})}\BibitemShut {NoStop}%
\bibitem [{\citenamefont {{Alam}}\ \emph {et~al.}(2015)\citenamefont {{Alam}},
  \citenamefont {{Sulaksono}},\ and\ \citenamefont {{Agrawal}}}]{Alam-15}%
  \BibitemOpen
  \bibfield  {author} {\bibinfo {author} {\bibfnamefont {N.}~\bibnamefont
  {{Alam}}}, \bibinfo {author} {\bibfnamefont {A.}~\bibnamefont {{Sulaksono}}},
  \ and\ \bibinfo {author} {\bibfnamefont {B.~K.}\ \bibnamefont {{Agrawal}}},\
  }\href {\doibase 10.1103/PhysRevC.92.015804} {\bibfield  {journal} {\bibinfo
  {journal} {\prc}\ }\textbf {\bibinfo {volume} {92}},\ \bibinfo {eid} {015804}
  (\bibinfo {year} {2015})}\BibitemShut {NoStop}%
\bibitem [{\citenamefont {{Kubis}}(2004)}]{Kubis-04}%
  \BibitemOpen
  \bibfield  {author} {\bibinfo {author} {\bibfnamefont {S.}~\bibnamefont
  {{Kubis}}},\ }\href {\doibase 10.1103/PhysRevC.70.065804} {\bibfield
  {journal} {\bibinfo  {journal} {\prc}\ }\textbf {\bibinfo {volume} {70}},\
  \bibinfo {eid} {065804} (\bibinfo {year} {2004})}\BibitemShut {NoStop}%
\bibitem [{\citenamefont {{Kubis}}(2007)}]{Kubis-07}%
  \BibitemOpen
  \bibfield  {author} {\bibinfo {author} {\bibfnamefont {S.}~\bibnamefont
  {{Kubis}}},\ }\href {\doibase 10.1103/PhysRevC.76.025801} {\bibfield
  {journal} {\bibinfo  {journal} {\prc}\ }\textbf {\bibinfo {volume} {76}},\
  \bibinfo {eid} {025801} (\bibinfo {year} {2007})}\BibitemShut {NoStop}%
\bibitem [{\citenamefont {Lattimer}\ and\ \citenamefont
  {Prakash}(2007)}]{Lattimer-07}%
  \BibitemOpen
  \bibfield  {author} {\bibinfo {author} {\bibfnamefont {J.~M.}\ \bibnamefont
  {Lattimer}}\ and\ \bibinfo {author} {\bibfnamefont {M.}~\bibnamefont
  {Prakash}},\ }\href@noop {} {\bibfield  {journal} {\bibinfo  {journal}
  {Phys.\ Rep.}\ }\textbf {\bibinfo {volume} {442}},\ \bibinfo {pages} {109}
  (\bibinfo {year} {2007})}\BibitemShut {NoStop}%
\bibitem [{\citenamefont {Nielsen}\ \emph {et~al.}(1991)\citenamefont
  {Nielsen}, \citenamefont {Provid{\^e}ncia},\ and\ \citenamefont
  {da~Provid{\^e}ncia}}]{Nielsen-91}%
  \BibitemOpen
  \bibfield  {author} {\bibinfo {author} {\bibfnamefont {M.}~\bibnamefont
  {Nielsen}}, \bibinfo {author} {\bibfnamefont {C.}~\bibnamefont
  {Provid{\^e}ncia}}, \ and\ \bibinfo {author} {\bibfnamefont {J.}~\bibnamefont
  {da~Provid{\^e}ncia}},\ }\href@noop {} {\bibfield  {journal} {\bibinfo
  {journal} {Phys.\ Rev.\ C}\ }\textbf {\bibinfo {volume} {44}},\ \bibinfo
  {pages} {209} (\bibinfo {year} {1991})}\BibitemShut {NoStop}%
\bibitem [{\citenamefont {Nielsen}\ \emph {et~al.}(1993)\citenamefont
  {Nielsen}, \citenamefont {Provid\^encia},\ and\ \citenamefont
  {da~Provid\^encia}}]{Nielsen-93}%
  \BibitemOpen
  \bibfield  {author} {\bibinfo {author} {\bibfnamefont {M.}~\bibnamefont
  {Nielsen}}, \bibinfo {author} {\bibfnamefont {C.}~\bibnamefont
  {Provid\^encia}}, \ and\ \bibinfo {author} {\bibfnamefont {J.}~\bibnamefont
  {da~Provid\^encia}},\ }\href@noop {} {\bibfield  {journal} {\bibinfo
  {journal} {Phys.\ Rev.\ C}\ }\textbf {\bibinfo {volume} {47}},\ \bibinfo
  {pages} {200} (\bibinfo {year} {1993})}\BibitemShut {NoStop}%
\bibitem [{\citenamefont {Avancini}\ \emph {et~al.}(2005)\citenamefont
  {Avancini}, \citenamefont {Brito}, \citenamefont {Menezes},\ and\
  \citenamefont {Provid\^encia}}]{Avancini-05}%
  \BibitemOpen
  \bibfield  {author} {\bibinfo {author} {\bibfnamefont {S.~S.}\ \bibnamefont
  {Avancini}}, \bibinfo {author} {\bibfnamefont {L.}~\bibnamefont {Brito}},
  \bibinfo {author} {\bibfnamefont {D.~P.}\ \bibnamefont {Menezes}}, \ and\
  \bibinfo {author} {\bibfnamefont {C.}~\bibnamefont {Provid\^encia}},\
  }\href@noop {} {\bibfield  {journal} {\bibinfo  {journal} {Phys.\ Rev.\ C}\
  }\textbf {\bibinfo {volume} {71}},\ \bibinfo {pages} {044323} (\bibinfo
  {year} {2005})}\BibitemShut {NoStop}%
\bibitem [{\citenamefont {Pais}\ and\ \citenamefont
  {Provid{\^e}ncia}(tion)}]{Pais-vlasov}%
  \BibitemOpen
  \bibfield  {author} {\bibinfo {author} {\bibfnamefont {H.}~\bibnamefont
  {Pais}}\ and\ \bibinfo {author} {\bibfnamefont {C.}~\bibnamefont
  {Provid{\^e}ncia}},\ }\href@noop {} {\  (\bibinfo {year} {in
  preparation})}\BibitemShut {NoStop}%
\bibitem [{\citenamefont {Provid\^encia}\ \emph
  {et~al.}(2006{\natexlab{b}})\citenamefont {Provid\^encia}, \citenamefont
  {Brito}, \citenamefont {Avancini}, \citenamefont {Menezes},\ and\
  \citenamefont {Chomaz}}]{ProvidenciaC-06}%
  \BibitemOpen
  \bibfield  {author} {\bibinfo {author} {\bibfnamefont {C.}~\bibnamefont
  {Provid\^encia}}, \bibinfo {author} {\bibfnamefont {L.}~\bibnamefont
  {Brito}}, \bibinfo {author} {\bibfnamefont {S.~S.}\ \bibnamefont {Avancini}},
  \bibinfo {author} {\bibfnamefont {D.~P.}\ \bibnamefont {Menezes}}, \ and\
  \bibinfo {author} {\bibfnamefont {P.}~\bibnamefont {Chomaz}},\ }\href@noop {}
  {\bibfield  {journal} {\bibinfo  {journal} {Phys.\ Rev.\ C}\ }\textbf
  {\bibinfo {volume} {73}},\ \bibinfo {pages} {025805} (\bibinfo {year}
  {2006}{\natexlab{b}})}\BibitemShut {NoStop}%
\bibitem [{\citenamefont {{Carriere}}\ \emph {et~al.}(2003)\citenamefont
  {{Carriere}}, \citenamefont {{Horowitz}},\ and\ \citenamefont
  {{Piekarewicz}}}]{Carriere03}%
  \BibitemOpen
  \bibfield  {author} {\bibinfo {author} {\bibfnamefont {J.}~\bibnamefont
  {{Carriere}}}, \bibinfo {author} {\bibfnamefont {C.~J.}\ \bibnamefont
  {{Horowitz}}}, \ and\ \bibinfo {author} {\bibfnamefont {J.}~\bibnamefont
  {{Piekarewicz}}},\ }\href {\doibase 10.1086/376515} {\bibfield  {journal}
  {\bibinfo  {journal} {\apj}\ }\textbf {\bibinfo {volume} {593}},\ \bibinfo
  {pages} {463} (\bibinfo {year} {2003})}\BibitemShut {NoStop}%
\bibitem [{\citenamefont {Ducoin}\ \emph
  {et~al.}(2008{\natexlab{b}})\citenamefont {Ducoin}, \citenamefont
  {Margueron},\ and\ \citenamefont {Chomaz}}]{Ducoin-08a}%
  \BibitemOpen
  \bibfield  {author} {\bibinfo {author} {\bibfnamefont {C.}~\bibnamefont
  {Ducoin}}, \bibinfo {author} {\bibfnamefont {J.}~\bibnamefont {Margueron}}, \
  and\ \bibinfo {author} {\bibfnamefont {P.}~\bibnamefont {Chomaz}},\
  }\href@noop {} {\bibfield  {journal} {\bibinfo  {journal} {Nucl.\ Phys.\ A}\
  }\textbf {\bibinfo {volume} {809}},\ \bibinfo {pages} {30} (\bibinfo {year}
  {2008}{\natexlab{b}})}\BibitemShut {NoStop}%
\bibitem [{\citenamefont {Horowitz}\ and\ \citenamefont
  {Shen}(2008)}]{Horowitz-08}%
  \BibitemOpen
  \bibfield  {author} {\bibinfo {author} {\bibfnamefont {C.~J.}\ \bibnamefont
  {Horowitz}}\ and\ \bibinfo {author} {\bibfnamefont {G.}~\bibnamefont
  {Shen}},\ }\href@noop {} {\bibfield  {journal} {\bibinfo  {journal} {Phys.\
  Rev.\ C}\ }\textbf {\bibinfo {volume} {78}},\ \bibinfo {pages} {015801}
  (\bibinfo {year} {2008})}\BibitemShut {NoStop}%
\bibitem [{\citenamefont {Sugahara}\ and\ \citenamefont
  {Toki}(1994)}]{Sugahara94}%
  \BibitemOpen
  \bibfield  {author} {\bibinfo {author} {\bibfnamefont {Y.}~\bibnamefont
  {Sugahara}}\ and\ \bibinfo {author} {\bibfnamefont {H.}~\bibnamefont
  {Toki}},\ }\href@noop {} {\bibfield  {journal} {\bibinfo  {journal} {Nucl.\
  Phys.\ A}\ }\textbf {\bibinfo {volume} {579}},\ \bibinfo {pages} {557}
  (\bibinfo {year} {1994})}\BibitemShut {NoStop}%
\bibitem [{\citenamefont {Todd-Rutel}\ and\ \citenamefont
  {Piekarewicz}(2005)}]{Todd05}%
  \BibitemOpen
  \bibfield  {author} {\bibinfo {author} {\bibfnamefont {B.~G.}\ \bibnamefont
  {Todd-Rutel}}\ and\ \bibinfo {author} {\bibfnamefont {J.}~\bibnamefont
  {Piekarewicz}},\ }\href@noop {} {\bibfield  {journal} {\bibinfo  {journal}
  {Phys.\ Rev.\ Lett.}\ }\textbf {\bibinfo {volume} {95}},\ \bibinfo {pages}
  {122501} (\bibinfo {year} {2005})}\BibitemShut {NoStop}%
\bibitem [{\citenamefont {Lalazissis}\ \emph {et~al.}(1997)\citenamefont
  {Lalazissis}, \citenamefont {K{\"o}nig},\ and\ \citenamefont
  {Ring}}]{Lalazissis-97}%
  \BibitemOpen
  \bibfield  {author} {\bibinfo {author} {\bibfnamefont {G.~A.}\ \bibnamefont
  {Lalazissis}}, \bibinfo {author} {\bibfnamefont {J.}~\bibnamefont
  {K{\"o}nig}}, \ and\ \bibinfo {author} {\bibfnamefont {P.}~\bibnamefont
  {Ring}},\ }\href@noop {} {\bibfield  {journal} {\bibinfo  {journal} {Phys.\
  Rev.\ C}\ }\textbf {\bibinfo {volume} {55}},\ \bibinfo {pages} {540}
  (\bibinfo {year} {1997})}\BibitemShut {NoStop}%
\bibitem [{\citenamefont {Moustakidis}\ \emph {et~al.}(2010)\citenamefont
  {Moustakidis}, \citenamefont {Niksic}, \citenamefont {Lalazissis},
  \citenamefont {Vretenar},\ and\ \citenamefont {Ring}}]{Moustakidis-10}%
  \BibitemOpen
  \bibfield  {author} {\bibinfo {author} {\bibfnamefont {C.~C.}\ \bibnamefont
  {Moustakidis}}, \bibinfo {author} {\bibfnamefont {T.}~\bibnamefont {Niksic}},
  \bibinfo {author} {\bibfnamefont {G.~A.}\ \bibnamefont {Lalazissis}},
  \bibinfo {author} {\bibfnamefont {D.}~\bibnamefont {Vretenar}}, \ and\
  \bibinfo {author} {\bibfnamefont {P.}~\bibnamefont {Ring}},\ }\href@noop {}
  {\bibfield  {journal} {\bibinfo  {journal} {Phys.\ Rev.\ C}\ }\textbf
  {\bibinfo {volume} {81}},\ \bibinfo {pages} {065803} (\bibinfo {year}
  {2010})}\BibitemShut {NoStop}%
\bibitem [{\citenamefont {Kr{\"u}ger}\ \emph {et~al.}(2013)\citenamefont
  {Kr{\"u}ger}, \citenamefont {Tews}, \citenamefont {Hebeler},\ and\
  \citenamefont {Schwenk}}]{KTHS2013}%
  \BibitemOpen
  \bibfield  {author} {\bibinfo {author} {\bibfnamefont {T.}~\bibnamefont
  {Kr{\"u}ger}}, \bibinfo {author} {\bibfnamefont {I.}~\bibnamefont {Tews}},
  \bibinfo {author} {\bibfnamefont {K.}~\bibnamefont {Hebeler}}, \ and\
  \bibinfo {author} {\bibfnamefont {A.}~\bibnamefont {Schwenk}},\ }\href@noop
  {} {\bibfield  {journal} {\bibinfo  {journal} {Phys.\ Rev.\ C}\ }\textbf
  {\bibinfo {volume} {88}},\ \bibinfo {pages} {025802} (\bibinfo {year}
  {2013})}\BibitemShut {NoStop}%
\bibitem [{\citenamefont {{Fattoyev}}\ and\ \citenamefont
  {{Piekarewicz}}(2010)}]{Fattoyev-10a}%
  \BibitemOpen
  \bibfield  {author} {\bibinfo {author} {\bibfnamefont {F.~J.}\ \bibnamefont
  {{Fattoyev}}}\ and\ \bibinfo {author} {\bibfnamefont {J.}~\bibnamefont
  {{Piekarewicz}}},\ }\href {\doibase 10.1103/PhysRevC.82.025810} {\bibfield
  {journal} {\bibinfo  {journal} {\prc}\ }\textbf {\bibinfo {volume} {82}},\
  \bibinfo {eid} {025810} (\bibinfo {year} {2010})}\BibitemShut {NoStop}%
\bibitem [{\citenamefont {{Baym}}\ \emph {et~al.}(1971)\citenamefont {{Baym}},
  \citenamefont {{Pethick}},\ and\ \citenamefont {{Sutherland}}}]{bps}%
  \BibitemOpen
  \bibfield  {author} {\bibinfo {author} {\bibfnamefont {G.}~\bibnamefont
  {{Baym}}}, \bibinfo {author} {\bibfnamefont {C.}~\bibnamefont {{Pethick}}}, \
  and\ \bibinfo {author} {\bibfnamefont {P.}~\bibnamefont {{Sutherland}}},\
  }\href {\doibase 10.1086/151216} {\bibfield  {journal} {\bibinfo  {journal}
  {\apj}\ }\textbf {\bibinfo {volume} {170}},\ \bibinfo {pages} {299} (\bibinfo
  {year} {1971})}\BibitemShut {NoStop}%
\bibitem [{\citenamefont {Grill}\ \emph {et~al.}(2014)\citenamefont {Grill},
  \citenamefont {Pais}, \citenamefont {Provid\^encia}, \citenamefont
  {Vida{\~n}a},\ and\ \citenamefont {Avancini}}]{Grill-14}%
  \BibitemOpen
  \bibfield  {author} {\bibinfo {author} {\bibfnamefont {F.}~\bibnamefont
  {Grill}}, \bibinfo {author} {\bibfnamefont {H.}~\bibnamefont {Pais}},
  \bibinfo {author} {\bibfnamefont {C.}~\bibnamefont {Provid\^encia}}, \bibinfo
  {author} {\bibfnamefont {I.}~\bibnamefont {Vida{\~n}a}}, \ and\ \bibinfo
  {author} {\bibfnamefont {S.~S.}\ \bibnamefont {Avancini}},\ }\href@noop {}
  {\bibfield  {journal} {\bibinfo  {journal} {Phys.\ Rev.\ C}\ }\textbf
  {\bibinfo {volume} {90}},\ \bibinfo {pages} {045803} (\bibinfo {year}
  {2014})}\BibitemShut {NoStop}%
\end{thebibliography}
\end{document}